\documentclass[twocolumn]{aa}
\pdfoutput=1
\usepackage{graphicx,epsfig,fancyhdr,psfig,rotating,amsmath,epsf,txfonts}
\usepackage{epstopdf}                     
\usepackage{amssymb}                    
\usepackage{color}                       
\usepackage{url}                         
\usepackage[toc,page]{appendix}
\usepackage{blindtext,capt-of}

\def\kms{\rm km\;s$^{-1}$}

\begin{document}

\title{Active region upflows: 1.\\ Multi-instrument observations}

\author{K. Vanninathan\inst{1},
        M.S. Madjarska\inst{2},
        K. Galsgaard\inst{3},
        Z. Huang\inst{4},
        J.G. Doyle\inst{2}}

\offprints{kamalam.vanninathan@uni-graz.at}
\institute{Institute of Physics/IGAM, University of Graz, 8010 Graz, Austria\\
			 \email{kamalam.vanninathan@uni-graz.at} 
	 \and
	 Armagh Observatory, College Hill, Armagh BT61 9DG, N.Ireland, UK
            \and
             Niels Bohr Institute, Geological Museum, {\O}stervoldgade 5-7, 1350 Copenhagen K, Denmark       
	\and
           Shandong Provincial Key Laboratory of Optical Astronomy and Solar-Terrestrial Environment, Institute of Space Sciences, Shandong University, Weihai, 264209 Shandong, China
           }
 \date{\today}

\abstract
{Upflows at the edges of active regions --- called `AR outflows', are studied using multi-instrument observations.}
{This study intends to provide the first direct observational evidence on whether chromospheric jets play an important role in furnishing mass that could sustain coronal upflows. The evolution of the photospheric magnetic field  associated with the footpoints of the upflow region and the plasma properties of active region upflows are investigated aiming at providing such information for benchmarking data-driven modelling of this solar feature that is presented in Galsgaard et al. (2015).}
{By spatially and temporally combining multi-instrument observations obtained with the \textit{Extreme-ultraviolet Imaging Spectrometer} on board the \textit{Hinode}, the \textit{Atmospheric Imaging Assembly} and the \textit{Helioseismic Magnetic Imager} instruments on board the \textit{Solar Dynamics Observatory} and the \textit{Interferometric BI-dimensional Spectro-polarimeter} installed at the National Solar Observatory, Sac Peak, we study the plasma parameters of the upflows and the impact of the chromosphere on active region upflows.}
{Our analysis shows that the studied active region upflow presents similarly to those studied previously, i.e. it displays blue-shifted emission of 5 -- 20\,\kms\ in Fe\,{\sc xii} and Fe\,{\sc xiii} and its average electron density is 1.8$\times$10$^9$\,cm$^{-3}$ at 1\,MK. The time variation of the density is obtained showing no significant change (in a 3$\sigma$ error). The plasma density along a single loop is calculated revealing a drop of 50\% over a distance of $\sim$20\,000\,km along the loop. We find a second velocity component in the  blue wing of the Fe~{\sc xii} and Fe\,{\sc xiii} lines at 105\,\kms\ reported only once before. For the first time we study the time evolution of this component  at high  cadence and find that it is persistent during the whole observing period of 3.5\,hours with variations of only $\pm$15\,\kms.  We also, for the first time, study the evolution of the photospheric magnetic field at high cadence and find  that magnetic flux diffusion is responsible for the formation of the upflow region. High cadence H$\alpha$ observations  are used to study the chromosphere at the footpoints of the upflow region. We find no significant jet-like (spicule/rapid blue excursion) activity to account for several hours/days of plasma upflow. The jet-like activity in this region is not continuous and blueward asymmetries are a bare minimum. Using an image enhancement technique for imaging and spectral data, we show that the coronal structures seen in the AIA~193\,\AA\ channel is comparable to the EIS~Fe\,{\sc xii} images, while images in the AIA~171\,\AA\ channel reveals additional loops that are a result of contribution from cooler emission to this channel.}
{Our results suggest that at chromospheric heights there are no signatures that support the possible contribution of spicules to active region upflows. We suggest that magnetic flux diffusion is responsible for the formation of the coronal upflows. The existence of two velocity components possibly indicate the presence of two different flows which are produced by two different physical mechanisms, e.g. magnetic reconnection  and pressure-driven.}

\authorrunning{K. Vanninathan et al.}
\titlerunning{Active region upflows}

\maketitle

\section{Introduction}
The possible contribution of the surroundings of active regions (ARs) to the slow solar wind was first pointed out by \cite{Kojima1999}. By comparing velocity distributions at 2.5 $R_{\odot}$ and potential field extrapolations based on Kitt Peak magnetograms, they found that compact low-speed wind regions are associated with large magnetic flux expansions adjacent to ARs. The first identification of `continuous intermittent flows' with velocities from 5 to 20\,\kms\ at the periphery of an AR was made by \cite{Winebarger2001} using \textit{Transition Region And Coronal Explorer} (TRACE) 171\,\AA\ data. By comparing with a quasi-static model, the authors suggested that this can only be explained by plasma flow from the chromosphere to the corona.  The identification of  AR outflows  was difficult, until the launch of \textit{Hinode}, because of the lack of suitable observations (spectral temperature coverage). \citet{Sakao2007}  `asserted that these observations are possibly the first identification
of outflowing solar wind material'  in the vicinity of ARs.  They analysed images from the \textit{X-ray Telescope} (XRT) on board \textit{Hinode} and measured upward Doppler velocities of $\sim$50\,\kms\ using the \textit{Extreme-ultraviolet Imaging Spectrometer} (EIS) in a Fe\,{\sc xii} line (no wavelength is given in the paper). \cite{DelZanna2008} gave a detailed report on AR upflows (blue-shifted emission). The author found Doppler velocities of 5 -- 10\,\kms\  in the EIS Fe\,{\sc xii}~195.12\,\AA\  line and 10 -- 30\,\kms\ in the Fe\,{\sc xv}~284.16\,\AA\ line, occurring in areas of weak emission `in a sharp boundary between low-lying hot 3\,MK loops and higher cooler 1\,MK loops'. \citet{Harra2008} also studied the blue-shifts in the coronal lines of Fe\,{\sc xii} and Fe\,{\sc xv}  and estimated line-of-sight (LOS) velocities of 20 -- 50\,\kms. We should note that no absolute calibration does exist for EIS data and only for certain type of observations near absolute values can be obtained \citep{Young2012}. \citet {Bryans2010}  reported the presence of upflows at the boundaries of ARs lasting for several days. The authors identified multiple blue wing components in the EIS Fe\,{\sc xii} and  Fe\,{\sc xiii} lines with velocities reaching up to 200\,\kms. Upflows were detected at greater heights in the solar atmosphere (between 1.5 and 2.3\,$R_{\odot}$)  in the H\,{\sc i} Ly\,${\alpha}$ line and the O\,{\sc vi} doublet  (at 1031.9\,\AA\ and 1037.6\,\AA) using data from the \textit{Ultra-Violet Coronagraph Spectrometer} on board the \textit{Solar and Heliospheric Observatory}  \citep{Zangrilli2012}. This further reinforced the idea that active regions could be a possible source of the slow solar wind. Several authors investigated the direct link of the AR upflows to the solar wind and supported the idea \citep{He2010, vanDriel-Gesztelyi2012, Culhane2014, Brooks2015}.

Numerous studies have been carried out to understand where these upflows originate and to determine how  the plasma proceeds. \cite{DelZanna2008}
 compared the AR outflows to the blue-shifts observed in the gradual phase of ribbon flares which are also located at the boundaries of downflowing plasma. Blue-shifts in  transition-region  and coronal lines during flares are commonly interpreted as a signature of chromospheric evaporation driven by reconnection. Following this, several authors speculated that the chromosphere could hold a possible explanation about the origin of these upflows. \citet{McIntosh2009} suggested that blueward asymmetries observed in the  Si\,{\sc vii} (log\,T$_{max}$ = 5.8\,K), Si\,{\sc x} (log\,T$_{max}$ = 6.1\,K) and Fe\,{\sc xiv} (log\,T$_{max}$ = 6.3\,K)  lines near ARs may be a result of spicular activity in the chromosphere. \citet{He2010} used Hinode \textit{Solar Optical Telescope} (SOT) data in Ca\,{\sc ii}~H to study the plage area underlying coronal outflows. The authors reported three jets that occurred in a small area over a period of one hour and `inferred that the intermittent coronal outflow might be caused by the chromospheric
jets'. \cite{Nishizuka2011} studied an AR outflow in EIS sit-and-stare data and found both continuous outflows and waves, which propagate from the
base of the outflow region.  From the similarities in the Fe\,{\sc xii} line profiles (blueward asymmetry and Doppler velocity) of the outflow region and a  jet observed in the same field-of-view (FOV) in EIS observations, they suggested that the flows and the waves `originate in unresolved explosive events in the lower atmosphere of the corona'. Wavelet analysis of X-ray emission at the edge of an active region done by \cite{Guo2010} indicated that the outflows are sporadic and quasi-periodic with periods of  5 -- 10\,min which led them to conclude that they might be a consequence of intermittent small scale magnetic reconnection in the chromosphere or the transition region.

\citet{Ugarte-Urra2011} found that intensity fluctuations observed in the Fe\,{\sc xii}~195.12\,\AA\ line within the low density outflow region are transient on timescales of 5\,min suggesting that they could be a result of transient events.  The comparison of the fluctuations at coronal and transition region heights indicated no relation between them. \citet{Warren2011}, using EIS data, showed that the AR edges are dominated by downflows in cooler lines (Fe\,{\sc viii} and  Si\,{\sc vii}, log\,T$_{max}$ = 5.8\,K) which show a bright  fan-like structure  and that the outflows occur only at higher temperature, i.e  from  Fe\,{\sc xi}  to Fe\,{\sc xiii}. The authors concluded that the morphology of the outflows and the fans are different. They also remark that the outflows are observed in regions where there is no emission in Si\,{\sc vii} and that the fans observed in imaging data are not directly related to the AR outflows as observed in spectral data. \cite{Brooks2012} studied the physical properties (temperature, emission measure, chemical composition) of the plasma that causes the blueward asymmetries in coronal lines and deduced that they are of coronal origin. Thus the chromospheric linkage of AR outflows was put to question.

 Despite the numerous studies  on AR outflows many questions remain unanswered. Blue-shifted emission in fan-like features at the edges of ARs are mostly considered as outflows that contribute to the slow solar wind. A study by \citet{Boutry2012} investigated two neighbouring  ARs and concluded that the upflow observed at the edge of one  is linked to a downflow into the other. The mass flow that enters one of the ARs was estimated as 18\% of the total upflow from the other AR. The authors concluded that AR upflows are either large-scale loops with mass  flow or open  magnetic structures where some of the slow solar wind originates. Their former explanation seems to have been ignored in most of the studies reported on the subject so far. 
 
 The suggestion that the emission at different temperatures is related to two populations of magnetic field lines at the edges of ARs, where open  field lines guide plasma upflows and closed downflows, is equally not yet understood. Red-shifted emission in fans at lower temperatures that have the same orientation as the outflow fans  is unlikely to only be interpreted as closed field flow and not, for instance, as downflow in the same region of the upflow. Another still to be answered question is whether the loops seen in TRACE  and \textit{Atmospheric Imaging Assembly} (AIA) observations are in the same positions where the blue-shifted emission in spectral coronal  data are observed. We also have very limited knowledge on the magnetic field evolution related to AR upflows with only one report on the subject \citep{Harra2010}. The role of the chromosphere in AR upflows remains under debate. 
 
 Our study does not aim at finding firm answers to all of the open questions. We investigate in great detail the chromospheric origin of AR upflows and provide essential new information for  benchmarking data-driven modelling (Galsgaard at al. 2015, hereafter Paper~II). Multi-instrument data from ground and space based instruments are analysed to study an upflow region from the chromosphere to the corona. Chromospheric imaging spectroscopy is used for the first time to investigate the contribution of chromospheric events to an AR upflow. The photospheric magnetic field evolution is analysed and the plasma properties of the AR upflow, both spatially and temporally, are derived. This information is further used to compare with observables derived from data-driven modelling 
 of the observed region. We use H$\alpha$ imaging and spectral data from the \textit{Interferometric BIdimensional Spectrometer} (IBIS) in combination with spectral data from the EIS/Hinode, images from the AIA and magnetograms from the \textit{Helioseismic and Magnetic Imager} (HMI) on board the Solar Dynamic Observatory (SDO) to study the origin of AR upflows.  The paper is organized as follows: In Section 2 we present  the observations, the results and discussion are given in Section 3 and in Section 4 we state our conclusions along with the summary.

\section{Observations and Data Analysis} 

The observations used  in the present  study were taken during a carefully planned  multi-instrument observing campaign in November 2010. Coordinated space-based, Hinode \citep{Kosugi2007} and SDO \citep{Pesnell2012}, as well as ground-based, IBIS (at National Solar Observatory, USA), observations were targeting NOAA 11123 as shown in Fig.~\ref{aia_fig}.

\subsection{Space observations: AIA/HMI/SDO \& EIS/Hinode}
Longitudinal magnetograms  from the HMI/SDO \citep{Scherrer2012} were used, although SOT data were also available. This choice was made  because of the requirement for a large FOV when performing the magnetic field extrapolation. The  data were obtained using the Fe\,{\sc i} 6173\,\AA\ line with a cadence of 45\,s. For the present analysis, data with a cadence of 15\,min extending over a  time period of 48\,hrs were used. The magnetograms were  corrected for the geometrical projection of the LOS magnetic flux \citep{Hagenaar2001} as the AR under study does not lie at the solar disc  center.  Images from the AIA/SDO \citep{Lemen2012} in the 1700\,\AA, 1600\,\AA, 171\,\AA\ and 193\,\AA\ were used after the standard AIA data processing.

\begin{figure}[!h]
\centering
\vspace{0cm}
\hspace{-1cm}
\includegraphics[width=10cm]{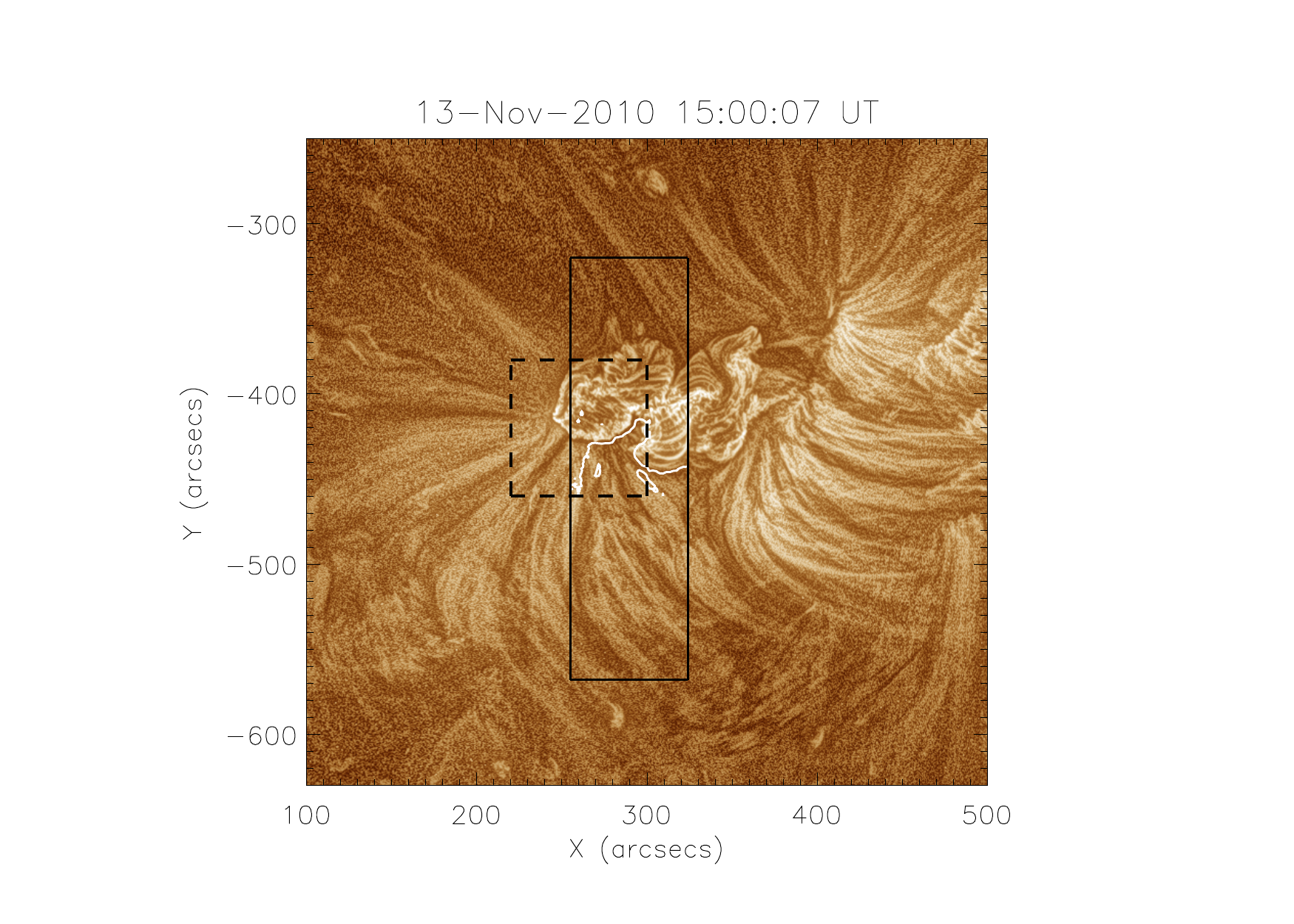}
\caption{AIA 193\,\AA\ MGN enhanced image of NOAA 11123. The solid black rectangle represents the EIS FOV for the large raster and the region enclosed by the  dashed line is the IBIS FOV. The white solid line is the contour of zero Doppler-shift outlining the upflow region. }
\label{aia_fig}
\end{figure}

The EIS \citep{Culhane2007} was observing between 15:02\,UT and 18:25\,UT on 2010 November 13.  The data
include one large raster ($70\arcsec \times248\arcsec$) (Fig.~\ref{aia_fig}) and 23 small rasters ($24\arcsec \times248\arcsec$) taken with 15\,sec exposure time and 75\,minutes of  sit-and-stare data with 5\,sec exposure time. Rotational compensation was applied during the entire observing period.  Several spectral lines were registered but for the present study we used the Fe\,{\sc xii}~195.12\,\AA\ and 186.88\,\AA, Fe\,{\sc xiii}~202.04\,\AA\ and 203.80\,\AA, Fe\,{\sc xiv}~274.20\,\AA\ and Fe\,{\sc xv}~284.16\,\AA\ lines.

In addition to the standard data processing, we also used the Multi-scale Gaussian Normalisation (MGN) code developed by \citet{Morgan2014}. This image processing permits the enhancement of fine structures in  high resolution data obtained from AIA/SDO and was  also successfully applied  to the EIS intensity images.

\subsection{Ground observations}
The IBIS instrument \citep{Cavallini2006, Reardon2008} recorded images in the H$\alpha$ 6562.8\,\AA\ line starting at 14:05\,UT. However, we did not use the first
2\,hrs of data because the seeing was variable. Between 15:50 and 16:30\,UT the seeing was most stable and further analysis was done using data from this time period. The IBIS FOV is shown in Fig.~\ref{aia_fig}. IBIS was setup to first scan the entire line by taking images at 15 positions along the line profile starting at 6561.47\,\AA\ (relative wavelength of $-$1.40\,\AA) in the blue wing and finishing at 6564.27\,\AA\ (relative wavelength of +1.40\,\AA) in the red wing. This was followed by  sets of 50 images at each of the wavelengths,  6562.07\,\AA\  (blue wing, relative wavelength of $-$0.80\,\AA),  6562.87\,\AA\ (line centre, relative wavelength of 0\,\AA) and at 6563.67\,\AA\  (red wing, relative wavelength of +0.80\,\AA). Each of these images were taken with 35\,ms exposure and this sequence was repeated for approximately three hours with minor breaks to resolve instrumental issues. Each set of 50 images was used to build a single speckle reconstructed image \citep{Woger2007}. The images in the line scans were not reconstructed and were used for line profile studies. The effective cadence of the dataset is 24\,s. Details of this observing campaign are given in \cite{Huang2014}.

\subsection{Data alignment}
All data analysed here were derotated to the reference time  15:00\,UT on 2010 November 13.  The AIA images were used as a reference to align the data from all other instruments. The AIA 1700\,\AA\ images were aligned with the IBIS H$\alpha$ blue wing images  and the data from the AIA 193\,\AA\ channel were used to align with the EIS big raster in the Fe\,{\sc xii}~195.12\,\AA\ line. The position of the EIS slit for the sit-and-stare observations was determined by comparing the intensities along the slit with the raster data. The HMI magnetograms were aligned with the imaging and the spectral data through the emission in the AIA 1600\,\AA\ channel. 

\section{Results and Discussion}
\subsection{EIS: Doppler-shifts and electron densities}

\begin{figure}[!h]
\centering
\vspace{0cm}
\hspace{-1cm}
\includegraphics[width=8cm]{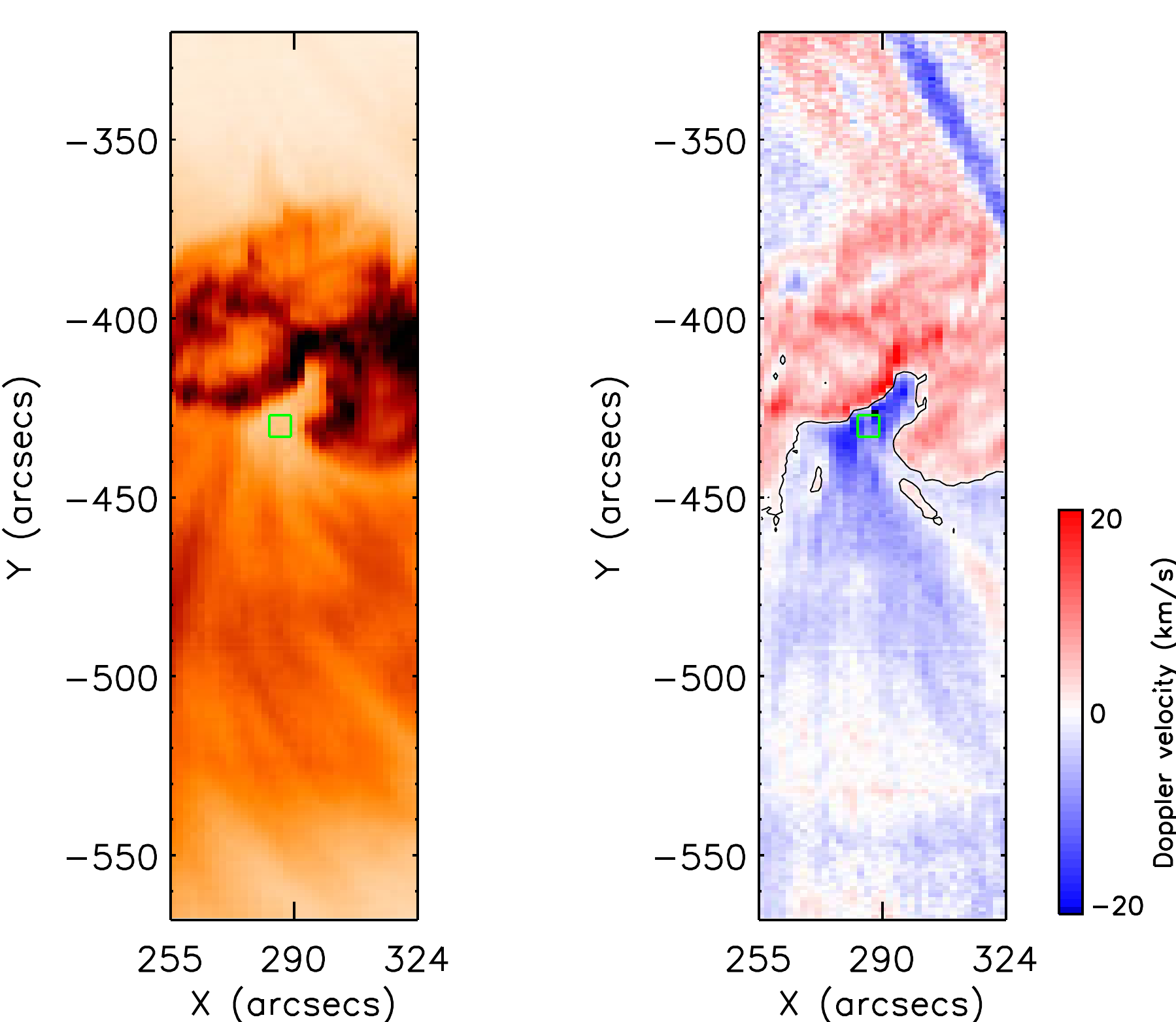}
\caption{EIS Fe\,{\sc xii}~195.12\,\AA\ intensity (\textbf{left}, in reverse colour table) and corresponding Doppler velocity (\textbf{right}) images taken starting at 15:02\,UT. The black contour line in the velocity image follows pixels with zero Doppler-shift. The green box indicates the region we have selected for further analysis.}
\label{eis_velocity}
\end{figure}

The initial aim of this study was to investigate the physical properties of the chromosphere and its impact on the upper solar atmosphere  using multi-instrument   data obtained during a dedicated observing campaign.  As we first produced the Doppler-shift images  in the EIS Fe\,{\sc xii}~195.12\,\AA\ line (Fig.~\ref{eis_velocity}), we found that a part of the observed AR exhibits plasma emission with blue-shifted Doppler velocities between 5 and 20$\pm$3\,\kms. The Doppler-shift image shown in Fig.~\ref{eis_velocity} was obtained by applying a single Gaussian fit. Due to the very limited FOV of the raster image covering entirely the AR, the rest component of the line was derived as an average over the entire FOV. The blue-shifted emission is at the edge of the AR and is associated with a region of reduced coronal emission (Fig.~\ref{eis_velocity}).  Thus, it has been identified as the phenomenon `AR outflow'. Note that we will mainly use the term upflow rather than outflow as it gives a more correct definition of the observed phenomenon. The blue-shifted emission along the LOS together with the fan-like orientation with respect to the observer  are consistent with the interpretation of an upflow. Further inspection of the small rasters and the sit-and-stare data (not shown here) revealed that the upflow lasted for the entire duration of the observing period, i.e. 3.5\,hrs. The EIS was also pointing at this region a day earlier, i.e. November 12, where the blue-shifted region from November 13 was not present. Nevertheless, another part of the active region showed a blue-shifted pattern. The blue-shifted region observed on November 13 (present data) appears to have been formed during the early hours of the same day, after a significant evolution of the entire region, and especially the upflow region. The evolution of the whole region as seen in the AIA~171\,\AA\ and 193\,\AA\ channels is shown in the online material (Figs.~\ref{fig_movie2} and \ref{fig_movie3}). We will come back to this while discussing the evolution of the magnetic field and coronal structures in Section 3.4.

We comment on the spatial offsets of the Doppler-shifts from regions where sharp intensity variation is observed as seen when comparing the intensity and velocity maps in  Fig.~\ref{eis_velocity}. \cite{Young2012} have pointed out that this offset is a result of distorted point-spread-function (PSF) that can introduce artificial velocity shifts in locations where there is a steep intensity gradient. The PSF effect appears to be present in our data between the bright closed loops and the low intensity upflow region producing a small offset in line-shift change boundary (from red to blue) with respect to the peak intensity. However, in the present analysis we are concerned with a large region displaying blue-shifts so the effect of the distorted PSF does not hinder our study. 

\begin{figure}[!h]
\centering
\vspace{0cm}
\hspace{-1cm}
\includegraphics[width=9cm]{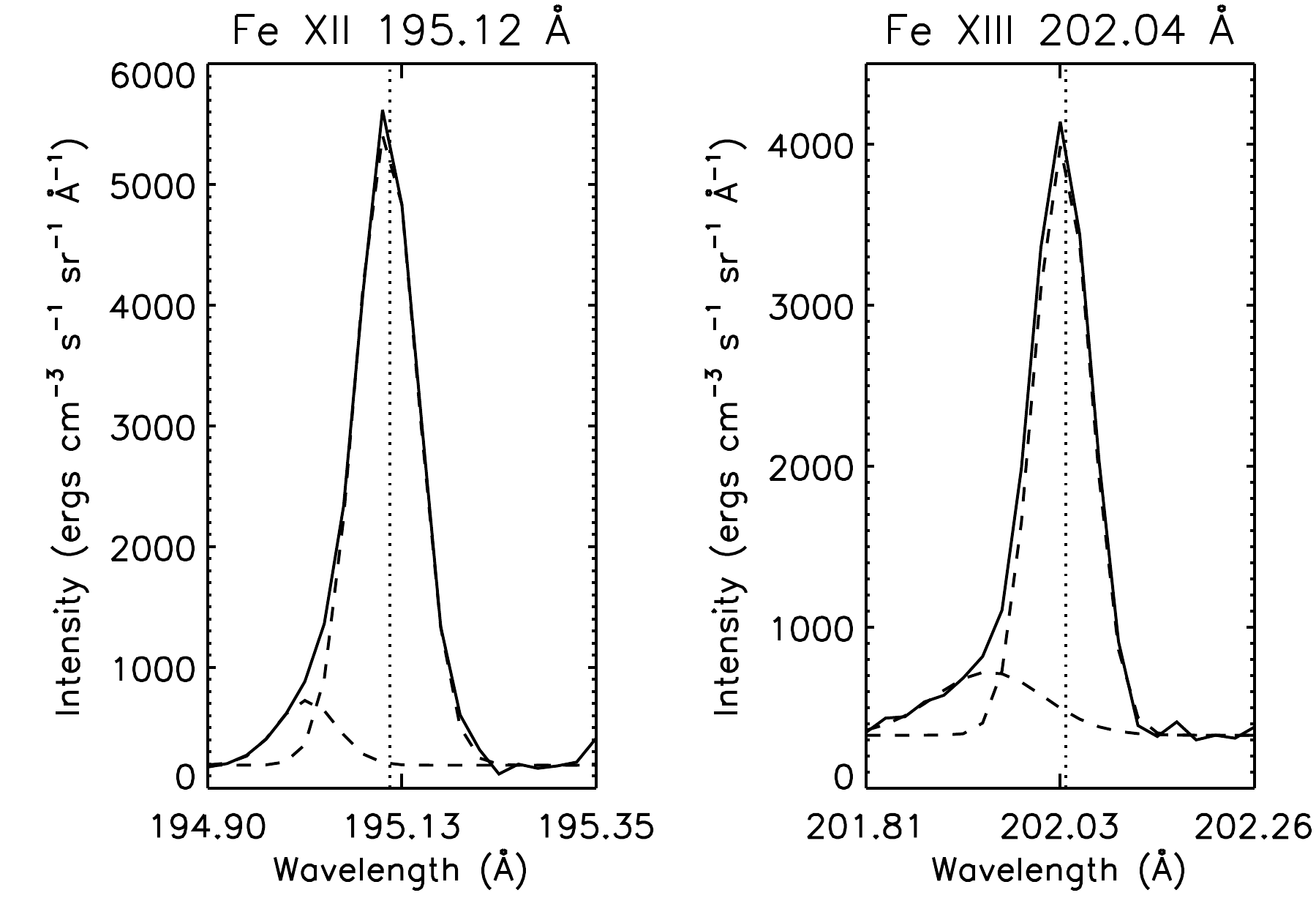}
\caption{EIS Fe\,{\sc xii}~195.12\,\AA\ (\textbf{left}) and Fe\,{\sc xiii}~202.04\,\AA\ ({\bf right}) lines at the footpoints of the upflow region taken from the boxed region shown in Fig.~\ref{eis_velocity}. The dashed lines in each of the panels represents the two fitted Gaussian components. The vertical dotted line shows the position of the rest component.}
\label{eis_asymmetry}
\end{figure}

\begin{figure}[!h]
\includegraphics[width=9cm]{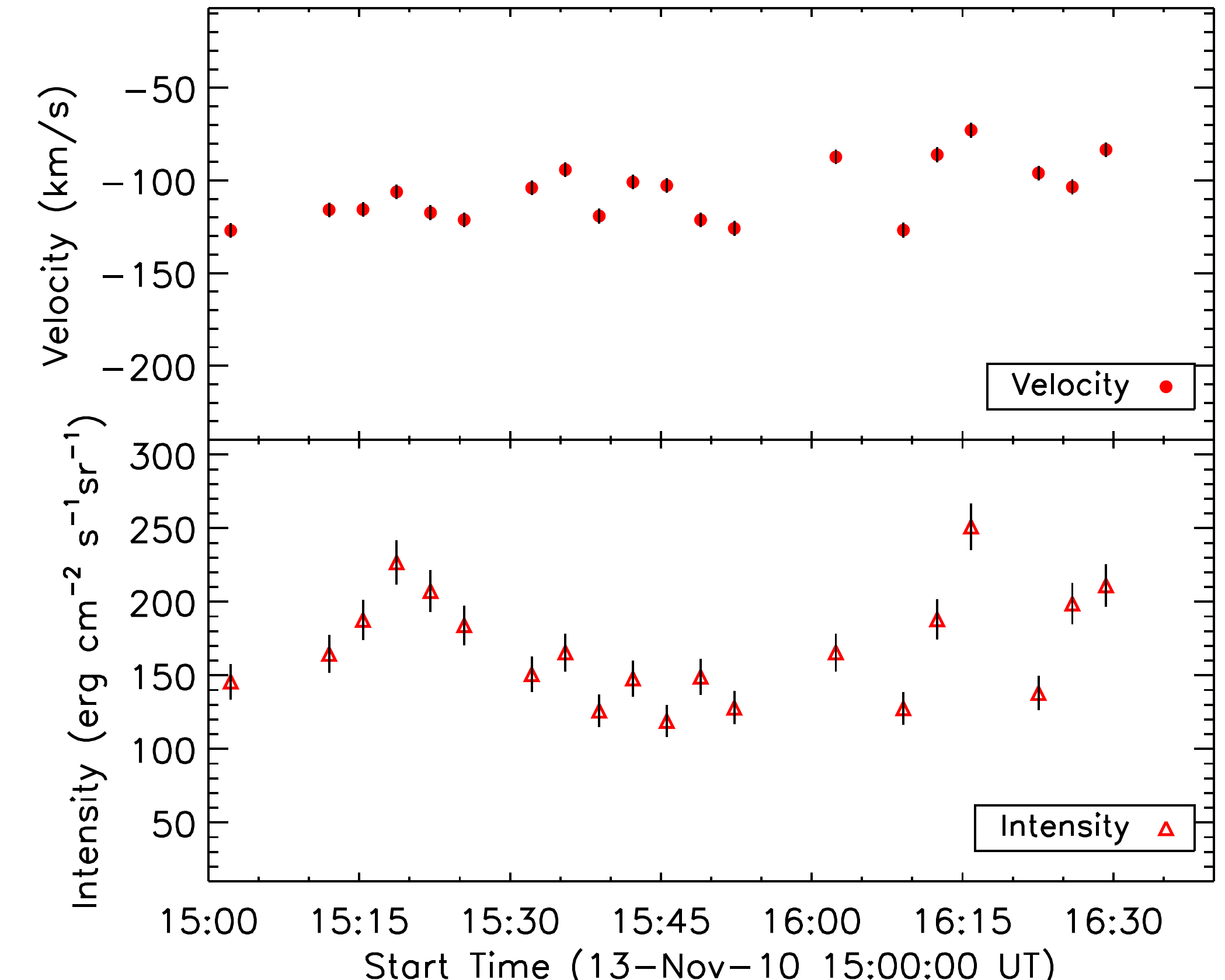}
\caption{Time evolution of the velocity with $\pm$4\kms\ error bars \textbf{(top)} and the intensity with errors \textbf{(bottom)} of the secondary component in the Fe\,{\sc xiii}~202.04~\AA\ line within the upflow region.}
\label{centroid}
\end{figure}

On account of the low exposure time used in this study (15\,s, note that the EIS study was not designed specifically for the investigation of AR outflows), binning of the photon signal from a small region close to the footpoints of the AR upflow was required. A box of size $6\arcsec \times7\arcsec$ (indicated in Fig.~\ref{eis_velocity}) was used for further analysis. We found that in this region there exists a secondary component in the blue wing of the spectral lines formed between 1 and 2\,MK (Fe\,{\sc xii}~195.12\,\AA\ and Fe\,{\sc xiii}~202.04\,\AA, see Fig.~\ref{eis_asymmetry}). This component was most pronounced in the Fe\,{\sc xiii} 202.04\,\AA\ line. It is difficult to confirm the presence of the secondary component in the Fe\,{\sc xiv} 274.20\,\AA\ line due to a blend with the Si\,{\sc vii} 274.18\,\AA\ line while the Fe\,{\sc xv}~284.16\,\AA\ line did not show any significant enhancements in the blue wing. This component was only seen close to the footpoints of the  upflow region. This result is consistent with the finding of \cite{Bryans2010}. In addition, we found that the emission in the second component decreased while moving away from the foopoints until it fully disappeared, while the primary component remained blue-shifted. The velocities derived from the Fe\,{\sc xiii} line for the secondary component close to the footpoints of the upflow region were of the order of 105\,\kms. Our dataset permitted us to study the time evolution of the second component for the first time. We found that it was persistent during the entire  observing period of 3.5\,hrs with a variation of $\pm$15\,\kms. These variations are relatively small compared to the absolute values (15\%). In Fig.~\ref{centroid} we have shown the variation of the centroid (top) and the variation in intensity (bottom) of the secondary component for the duration of our raster observations.

We did not observe any jet-like velocity or intensity structures in the EIS data within the upflow region although other related studies have reported the presence of such jets in the vicinity of the AR upflows \citep{Nishizuka2011}. The formation of coronal jets in ARs is a common phenomenon. However, they provide a much more localised flow profile \citep{Madjarska2011a}, and are unlikely to create a broad upflow region with a smooth flow profile over time. These jets need to occur at a constant rate to be able to sustain long duration upflows. Their occurrence will trigger significant magnetic field reconfiguration \citep{Moreno2013} which has not been observed so far. Hence, coronal jets cannot be responsible for the formation of AR upflows.

For the purpose of realistic modelling of solar phenomena, it is important to reach a deeper understanding  of the physical processes that take place in magnetised plasmas. We have made the first challenging  step towards combining solar observations and a numerical 3D magneto-hydrodynamic (MHD) experiment in a
so-called data-driven model (Paper~II). For this we use the present data to obtain all physical parameters together with the 
\begin{figure}[!h]
\vspace{-0.5cm}
\includegraphics[width=9cm]{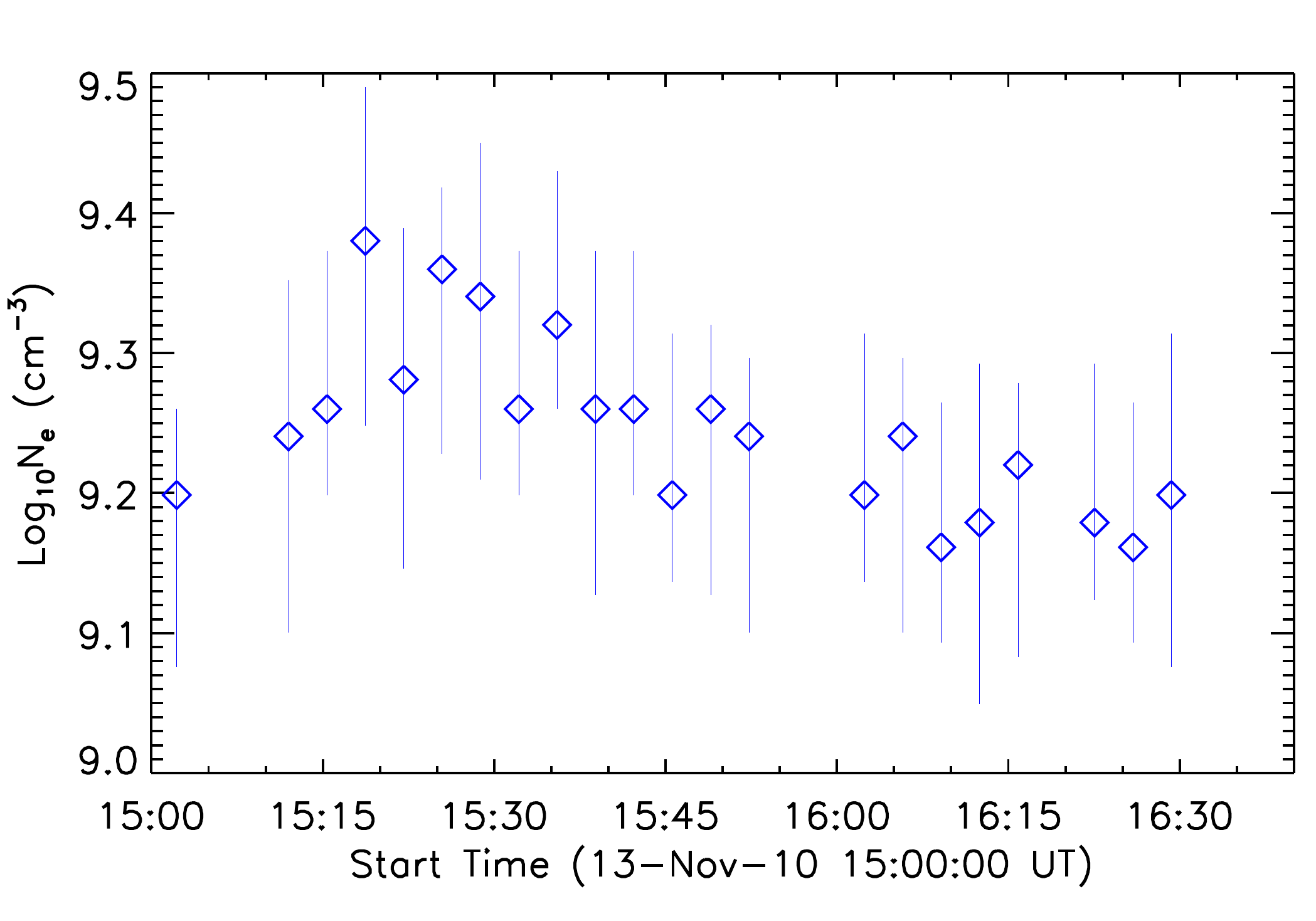}
\caption{Time evolution of the electron density within the upflow region calculated from the  Fe\,{\sc xii} line ratio with 3$\sigma$ errors plotted.}
\label{density_graph}
\end{figure}
magnetic field and coronal evolution that can be later used for comparison with the observables  derived from the model. 

We calculated the electron densities close to the footpoint of the upflow region using the density sensitive  line ratio 
of  Fe\,{\sc xii} computed as (186.88\,\AA\ + 186.85\,\AA)/(195.12\,\AA\  + 195.18\,\AA). The measured density in the region enclosed within the box  shown in Fig.~\ref{eis_velocity} is 1.58$\times10^{9}$\,cm$^{-3}$. We did not use the Fe\,{\sc xiii} line intensity ratio, (203.82\,\AA\ + 203.79\,\AA)/(202.04\,\AA), because, a blend by Fe\,{\sc xii}~203.72\,\AA\ in the wing of Fe\,{\sc xiii}~203.82\,\AA\ is impossible to remove as it is merged with the second velocity component explained above. Since space and time variations of physical parameters give important clues about the physical process at work, we also derived the time variation of the electron density using the data from the large and small rasters.
The electron density showed no significant changes during the rastering-mode observing period (Fig.~\ref{density_graph}). Since binning over a small region is required, we limited our time variation analysis to the raster data and did not make use of the sit-and-stare data. 

\begin{figure}[htp!]
\vspace{-1cm}
\begin{minipage}[b]{0.6\linewidth}
\includegraphics[width=5cm]{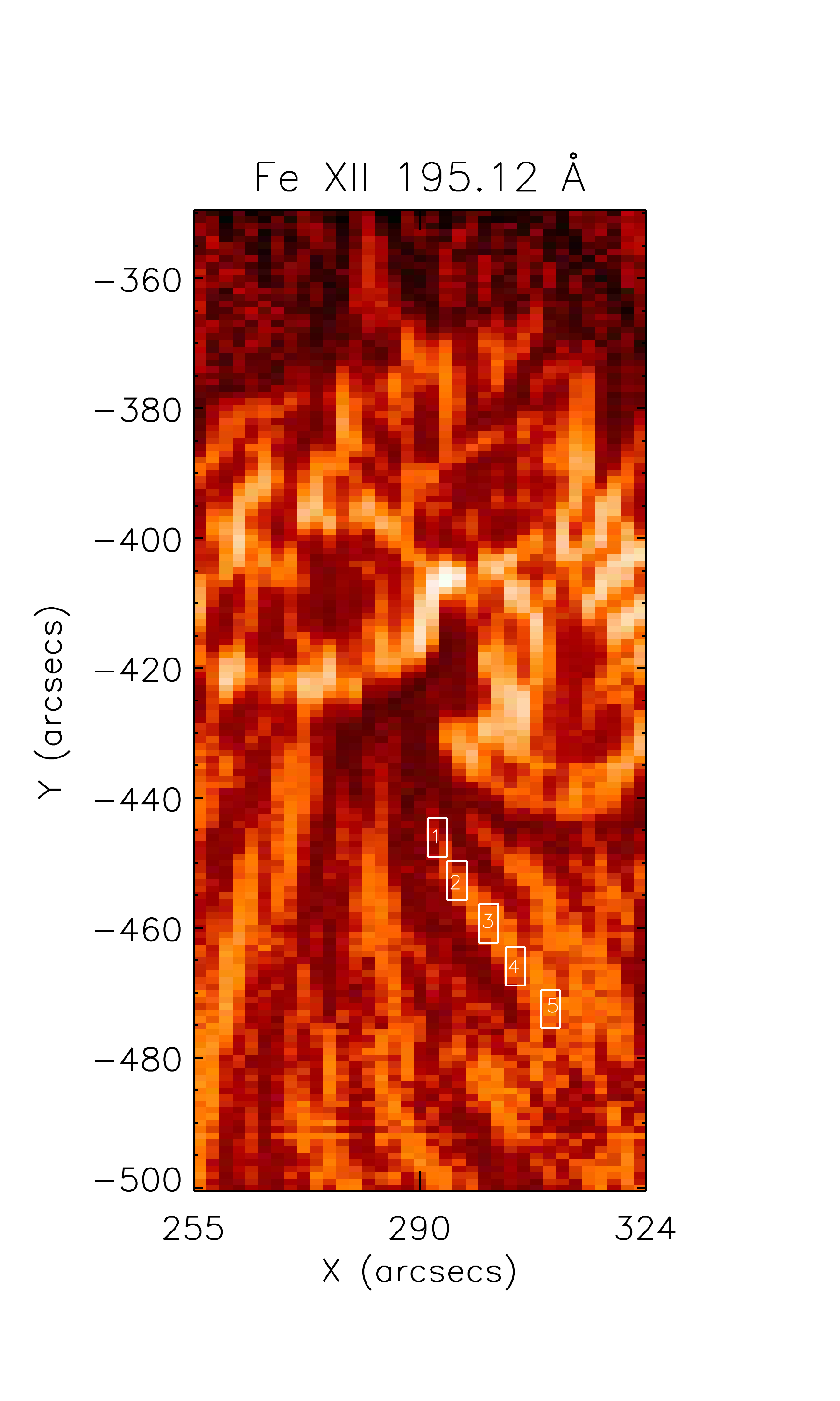}
\end{minipage}
    \hspace{-1cm}
\begin{minipage}[b]{0.3\linewidth}
    \begin{tabular}{ccc}
        \hline
        \hline
        Index & \multicolumn{1}{c}{N$_e$ ($\times10^9$\,cm$^{-3}$)} \\ 
        & Fe\,{\sc xii} \\
        \hline 
        \hline 
        1 & 0.79 \\
        2 & 0.76 \\
        3 & 0.72 \\
        4 & 0.58  \\
        5 & 0.36 \\
        \hline
      \end{tabular}
      \label{density_table}
      \par\vspace{1cm}
\end{minipage}
\caption{{\bf Left:} MGN enhanced EIS Fe\,{\sc xii}~195.12\,\AA\ image of the upflow region. Boxes are marked along one of the quasi-open loops. {\bf Right:} Densities of the boxed regions as calculated from the Fe\,{\sc xii} line ratio.}
\label{loop_density}
\end{figure}

To compare with the simulations done in Paper~II, we derived the electron densities along one of the quasi-open loops from the boxed regions shown in Fig.~\ref{loop_density} (left), using the Fe\,{\sc xii} line ratio. The obtained densities are given in the table in  Fig.~\ref{loop_density} (right). The AIA data have better spatial resolution as well as higher signal in comparison to the present EIS dataset. By comparing the AIA 171~\AA\ and 193~\AA\ images, we were able to follow the quasi-open loop further out than that visible in EIS data. It is seen to extend up to y=$-$425\arcsec\ which  is assumed to be the footpoint of this feature.
 Please note that we are mainly interested in the relative distance from box 1 to box 5 rather then the precise position of the footpoint of the loop.
We calculated linear distances along this loop. We did not take into account the projection effect and assumed that the loop is straight. Since 1\arcsec\ on the Sun is equivalent to 750\,km, we calculated the distance from the base of the loop to region 1 to be 16\,500\,km. From region 1 to region 5 (which is at a distance of 36\,000\,km) we found that the density  falls by $\sim$50\%. We further discuss these measurements in Paper~II.

\subsection{Magnetic field evolution}
\begin{figure}[ht!]
\centering
\vspace{-2cm}
\includegraphics[width=9cm]{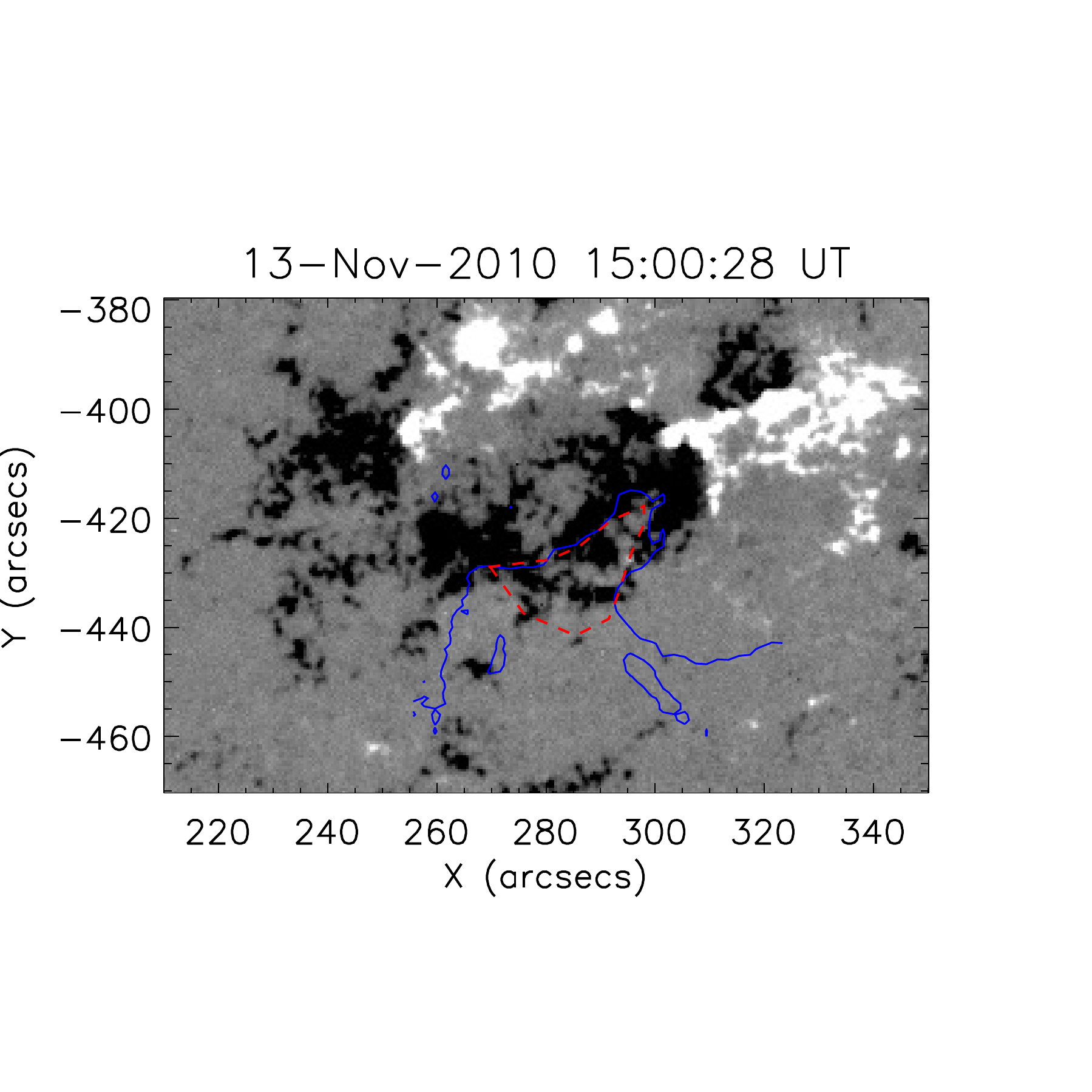}
\vspace{-1.5cm}
\caption{HMI magnetogram taken on 13 November 2010 at 15:00\,UT with the contours of the EIS upflow region derived from the EIS big raster (shown in Fig.~\ref{eis_velocity}, right) marked in solid blue. The area bounded by the red dashed line is the footpoints of the upflows and was used for the study of the magnetic flux evolution shown in Fig.~\ref{lightcurve}.}
\label{unipolar}
\end{figure}

\begin{figure}[ht!]
\centering
\includegraphics[width=8cm]{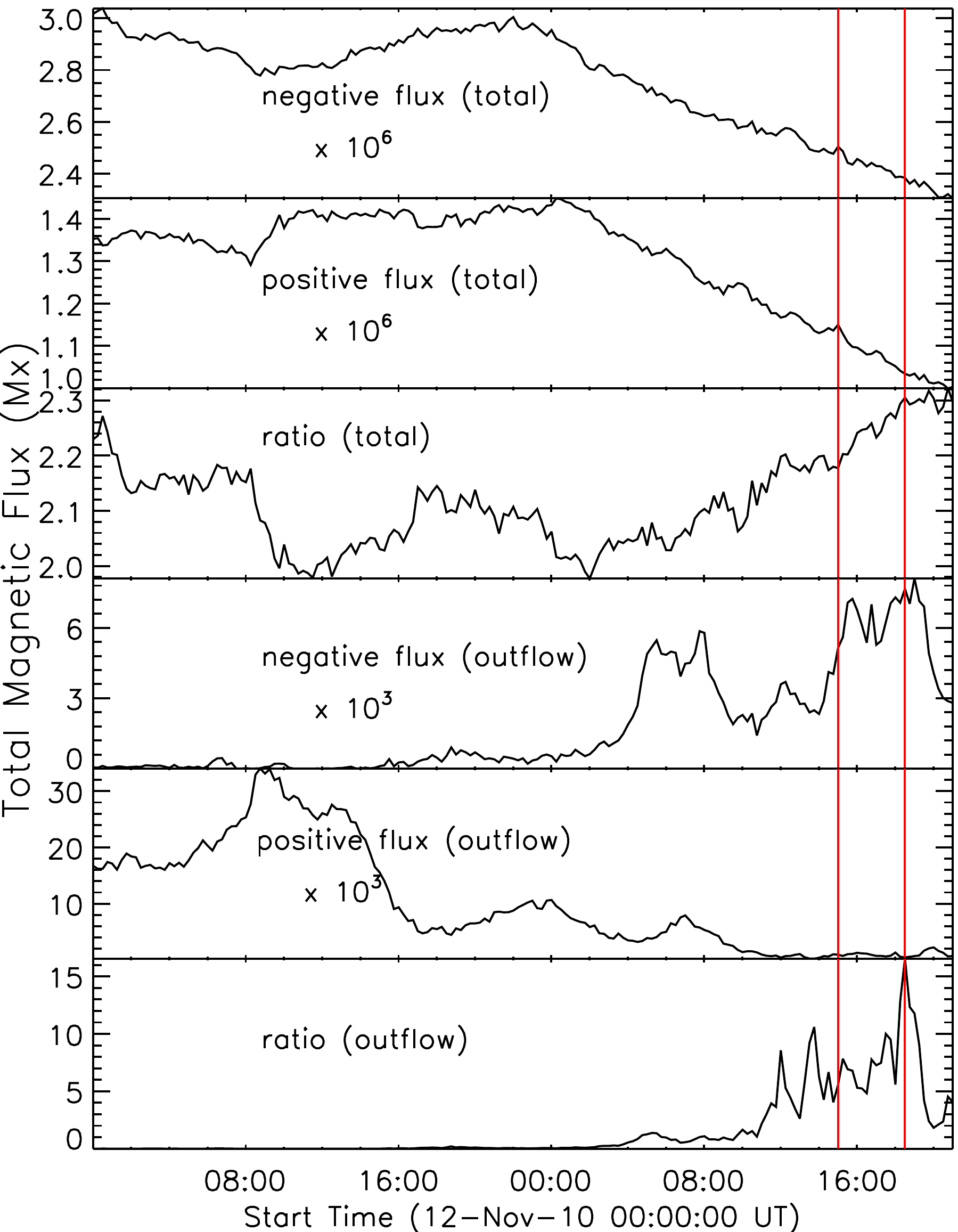}
\caption{Time evolution of the absolute magnetic flux of each polarity within the full FOV (top 3 panels) shown in Fig.~\ref{unipolar} and within the region enclosed by the red dashed line (bottom 3 panels) in the same figure. The vertical  lines denote the period during which EIS observations are available.}
\label{lightcurve}
\end{figure}

The next logical step of our analysis was to extract all the necessary information from the magnetic field and coronal imaging data that could be  of importance to the modelling part of this study. After aligning the data from the three instruments, EIS, HMI and AIA, we combined them in an image sequence shown as animations in the online material (Figs.~\ref{fig_movie2} and \ref{fig_movie3}). The animations clearly show the roots of the `loops'  in the photosphere along where the upflow occurs and how they evolve in time. Thus the footpoint of the upflow region was  manually  selected  by following the EIS velocity contour while cross checking with the combined HMI and AIA imaging data.  Fig.~\ref{unipolar} shows that the observed blue-shifted fan-like feature is rooted in  a single polarity of the AR. The same has been previously reported for two ARs that were studied by \cite{Doschek2008} and \cite{Baker2009}. We  used the very basic approach of producing lightcurves to study the general evolution of the magnetic field in the AR.      
The lightcurves of the magnetic flux (Fig.~\ref{lightcurve}) were  derived for the full FOV of Fig.~\ref{unipolar} and for the footpoint of the upflows enclosed within the red dashed line. We used data starting on November 12, 00:00 UT and lasting for 48 hours with a cadence of 15\,min. This period includes the time during which the EIS observations were obtained  (from 15:02\,UT to 18:30\,UT, indicated by vertical lines in Fig.~\ref{lightcurve}). We found that there was no significant change in the overall magnetic field of the AR (FOV shown in Fig.~\ref{unipolar}) until 00:00\,UT on November 13 (see top 3 panels of Fig.~\ref{lightcurve}). After this time there is a continuous decrease of both the negative and the positive fluxes at rates of  1.04$\times 10^{16}$\,Mx s$^{-1}$ and 7.54$\times 10^{15}$\,Mx s$^{-1}$, respectively. The ratio of the two opposite fluxes indicates an increase in the flux imbalance from 2.0 to 2.3, i.e by 15\%. The lightcurve of the footpoints of the upflow region clearly shows that the negative flux increased by a factor of 8 from 00:00\,UT to 24:00\,UT on November 13,  while the positive flux went down almost 8 times after 07:00\,UT on November 13 (see bottom 3 panels of Fig.~\ref{lightcurve}). We made an additional  visual inspection and found two distinct changes. First, a small bipolar region that was partially in the footpoints of the upflow region, disappeared after flux cancellation with almost full  elimination of the positive flux which explains the strong decrease of the positive flux in the upflow region.  Second, negative flux diffused into the region of the upflow footpoints. This evolution can be followed in the online material (Figs.~\ref{fig_movie2} and \ref{fig_movie3}). The resulting magnetic flux ratio clearly indicates a picture of a region dominated by a single polarity. Through observations and 3D MHD simulations \cite{Harra2010} and \cite{Harra2012}, respectively, found  that the edges of flux emerging regions show distinct blue-shifts which could result from either reconnection jets or pressure driven jets. In the present case flux diffusion along with a complex evolution of the AR (which cannot be understood by simple lightcurves of the magnetic flux) rather than flux emergence is responsible for the formation of the AR upflow. 

\subsection{IBIS}
 The nature of the upflow --- a steady continuous plasma flow with very little or no variation in time, suggests that chromospheric phenomena such as high velocity plasma ejections called spicules may be a suitable candidate. Such chromospheric small-scale ejections could be masked by the low spatial resolution of EIS and spatially appear as unresolved large-scale flows. Due to the frequency of occurrence of such jets and their ubiquitousness, the chromospheric origin of the AR upflows has been proposed \citep{McIntosh2009, He2010, Nishizuka2011}. The IBIS data provide a unique opportunity for exploring this idea. We used high cadence (24\,s) spectral imaging data taken in the H$\alpha$ line. These data are perfectly suitable for studying chromospheric ejections like type~I and II spicules, rapid blue excursions (RBEs), fibrils, mottles etc. \citep[for a review see][]{Tsiropoula2012}. 
 
 The IBIS images were overlayed with the EIS contour of the blue-shifted coronal emission to determine the region associated with the AR upflow. The footpoints of the upflow were visually inspected for the presence  of chromospheric jet-like features. We found no evidence for the presence of such phenomena in the abundance required for sustaining a coronal plasma upflow for several hours or days. As mentioned above, type~II spicules were proposed as a candidate to sustain the continuous plasma upflow at the edges of ARs \citep{McIntosh2009}. Type~II spicules show upward motion with no indication of downward motion which has been interpreted as mass deposition in the upper solar atmosphere.  Although the coronal upflow lasted for several hours, we found that the footpoints of the upflow in the chromosphere did not contain ample number of jets (type~II spicules or otherwise) which are capable of feeding such an upflow. 

\begin{figure*}[!ht]
\centering
\vspace{-2.5cm}
\includegraphics[scale=0.8]{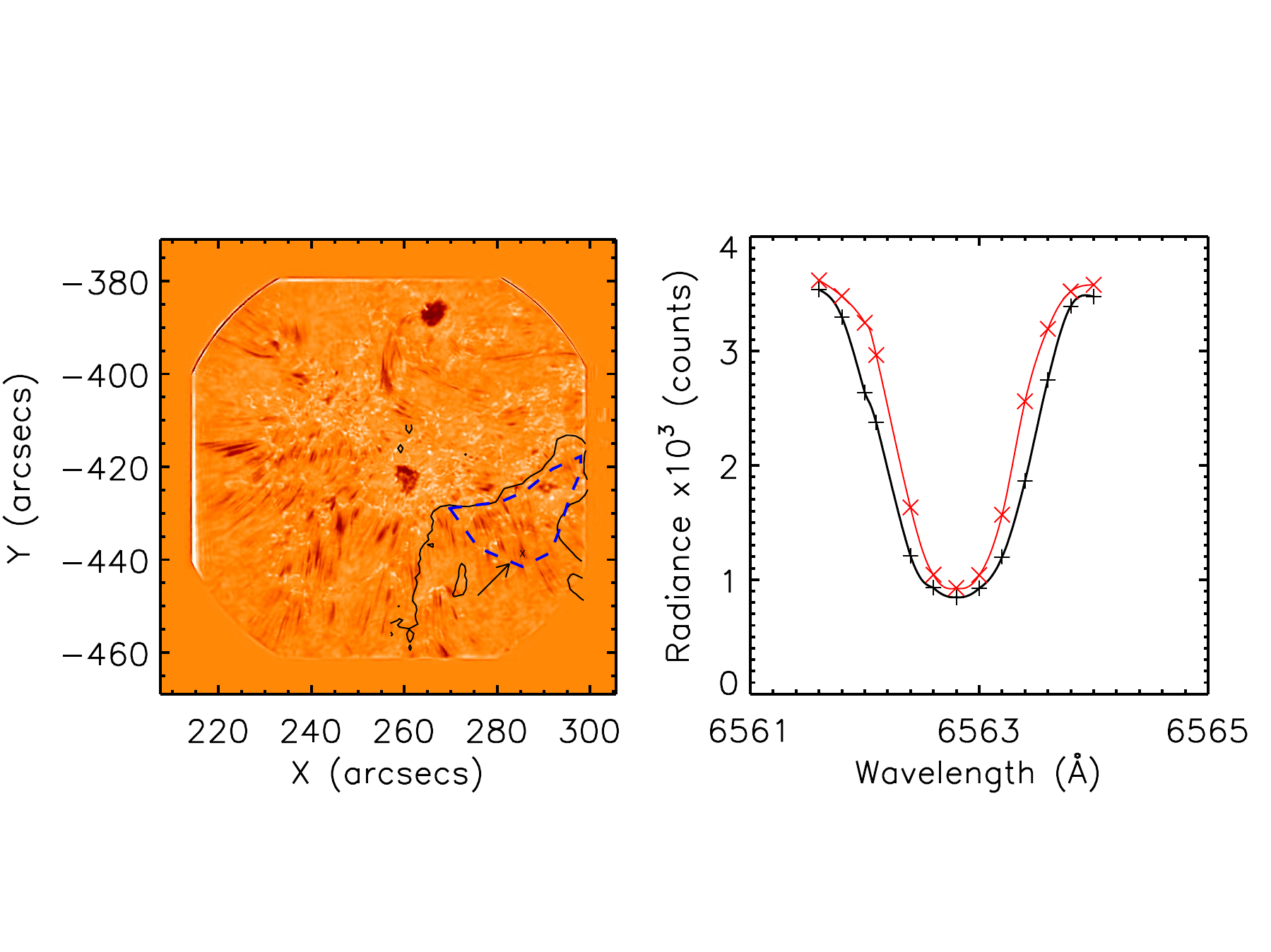}
\vspace{-1.5cm}
\caption{{\bf Left:} Full FOV of a speckle reconstructed IBIS image at $-$0.8\,\AA\ ($-$36\,\kms). The contour of the upflow region as derived from the EIS Doppler-shift image is superimposed in black. The dashed blue line is the contour of the footpoints of the upflow region. {\bf Right:} Average H$\alpha$ line profile in red and in black the line profile of a single pixel in a jet (spicule or RBE) within the footpoints of the upflow region. The pixel is marked by a cross in the left panel and pointed at with an arrow.}
\label{blue_compare}
\end{figure*}

A statistical study on the properties of RBEs (believed to represent the on-disk counterpart of type~II spicules) by \cite{Rouppe2009} suggested that these jets have an average blue-shifted Doppler velocity of 35\,\kms\ in H$\alpha$ data. This corresponds to $\Delta\lambda$ = $-$0.8\,\AA\ which is the same wavelength position at which the current data were obtained.  The authors also identify RBEs as isolated dark streaks at wavelength positions corresponding to $-$59.4\,\kms, $-$45\,\kms\ and $-$73.1\,\kms\ in the H$\alpha$ line.
In Fig.~\ref{blue_compare} (left)  we show the jets  at a wavelength position of $\Delta\lambda$ = $-$0.8\,\AA\ which
corresponds to Doppler shifts of 36\,\kms.  From this figure we see that, although there are a few jets with velocities of $-$36\,\kms, the line profile does not indicate excess velocity in the blue wing which is considered a characteristic of RBEs. This throws  further doubts on the chromospheric origin of AR upflows.  

\begin{figure}
\centering
\includegraphics[width=7cm]{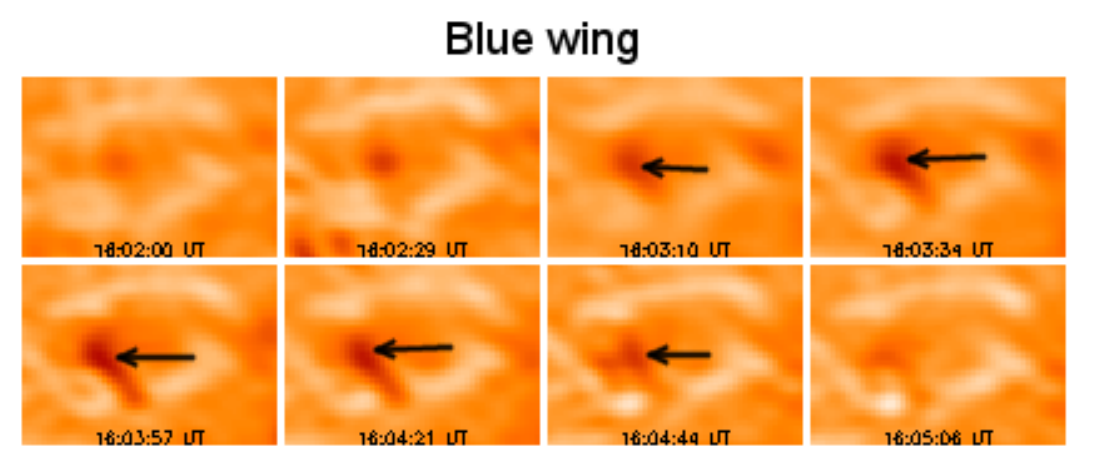}
\includegraphics[width=7cm]{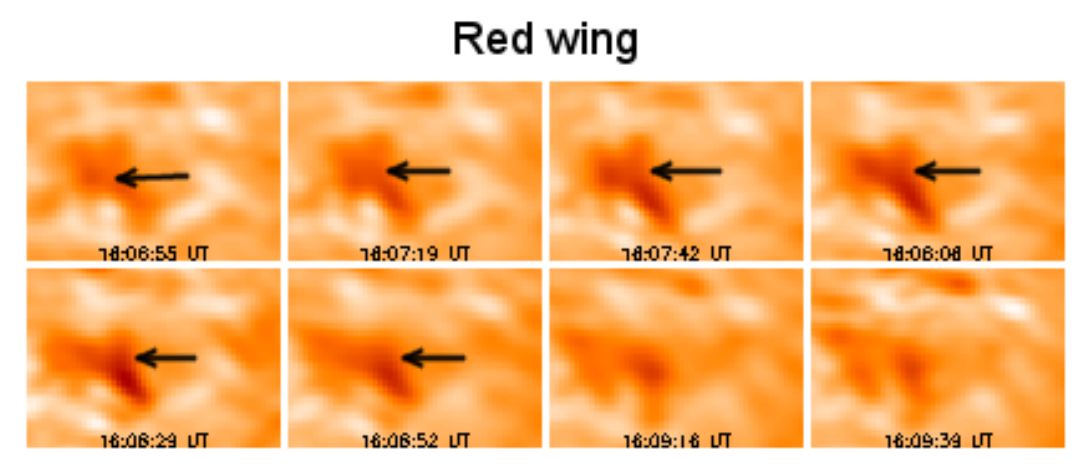}
\includegraphics[width=7cm]{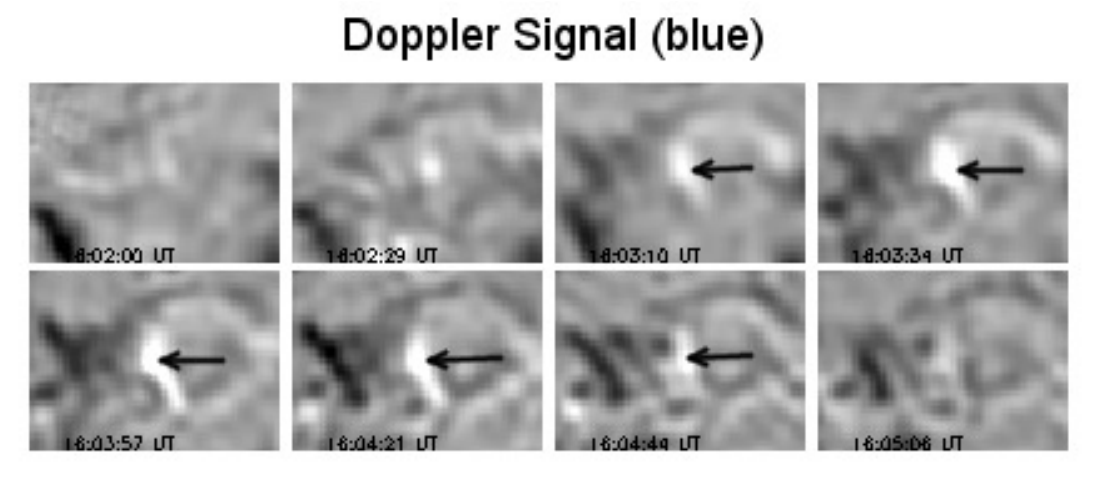}
\includegraphics[width=7cm]{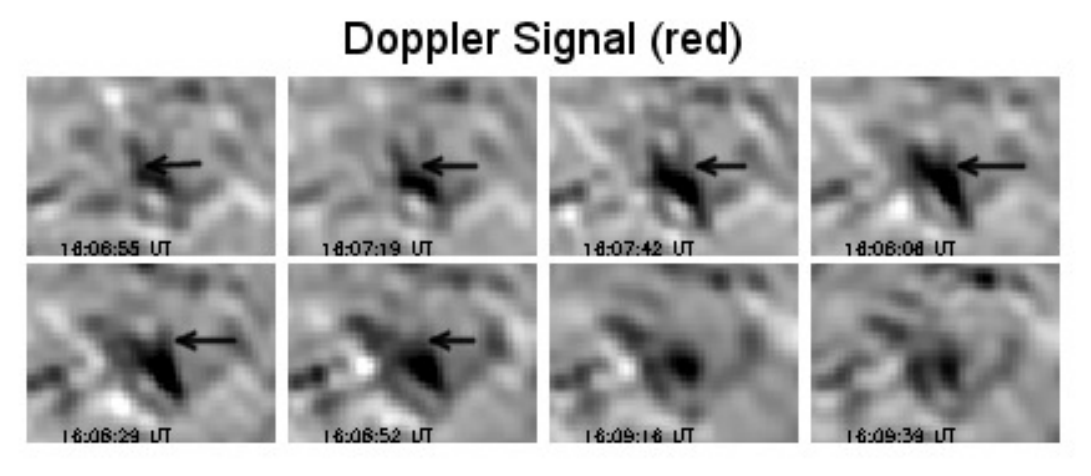}
\caption{Time evolution of a jet (indicated by an arrow in each panel) in the blue wing of the H$\alpha$ line (top panels) and in the red wing (second set of panels) and the corresponding DS images (bottom two sets of panels. }
\label{jet evolution}
\end{figure}

We made a more detailed analysis by randomly studying the evolution of jets in the footpoints of the upflow region in both the blue and the red wings of the H$\alpha$ line. We found that they show both plasma upflows as well as downflows which is not expected from type~II spicules. An example of one such jet is shown in Fig.~\ref{jet evolution}. In this example we found that the jet first showed absorption in the blue wing starting at 16:02:00\,UT (first set of panels labeled `Blue wing'). This absorption  reached a maximum around 16:03:57\,UT and then slowly faded away 1\,min later at around 16:05:06\,UT. To better illustrate the dynamics that takes place, we also made Doppler signal (DS) images using the formula 

\begin{equation}
\vspace{-1cm}
DS~=~\frac{I(+\Delta\lambda)~-~I(-\Delta\lambda)}{I(+\Delta\lambda)~+~I(-\Delta\lambda)},
\end{equation}
where, $+\Delta\lambda$ is $+$0.8\,\AA\ (6563.67\,\AA) and $-\Delta\lambda$ is $-$0.8\,\AA\ (6562.07\,\AA). A positive DS denotes an upward motion of absorbing material.

From Fig.~\ref{jet evolution} (third set of panels labeled `Doppler Signal (blue)') we see that the increase in intensity in the blue wing corresponds to positive DS indicating
plasma moving up. Starting at 16:06:55\,UT we observed an increased emission in the red wing ($\Delta\lambda$ = $+$0.8\,\AA, see Fig.~\ref{jet evolution}, second set of panels labeled `Red wing') at the same location where we earlier saw increased emission in the blue wing images. The intensity increased until 16:08:06\,UT and then faded away by 16:09:39\,UT. The corresponding DS images (Fig.~\ref{jet evolution}, fourth set of panels labeled `Doppler Signal (red)') show negative signal which is representative of the plasma falling
back again.  Since we observe that the plasma returns to the chromosphere there is a doubt if enough material has been deposited in the solar corona. 

\begin{figure}[!h]
\centering
\includegraphics[width=8cm]{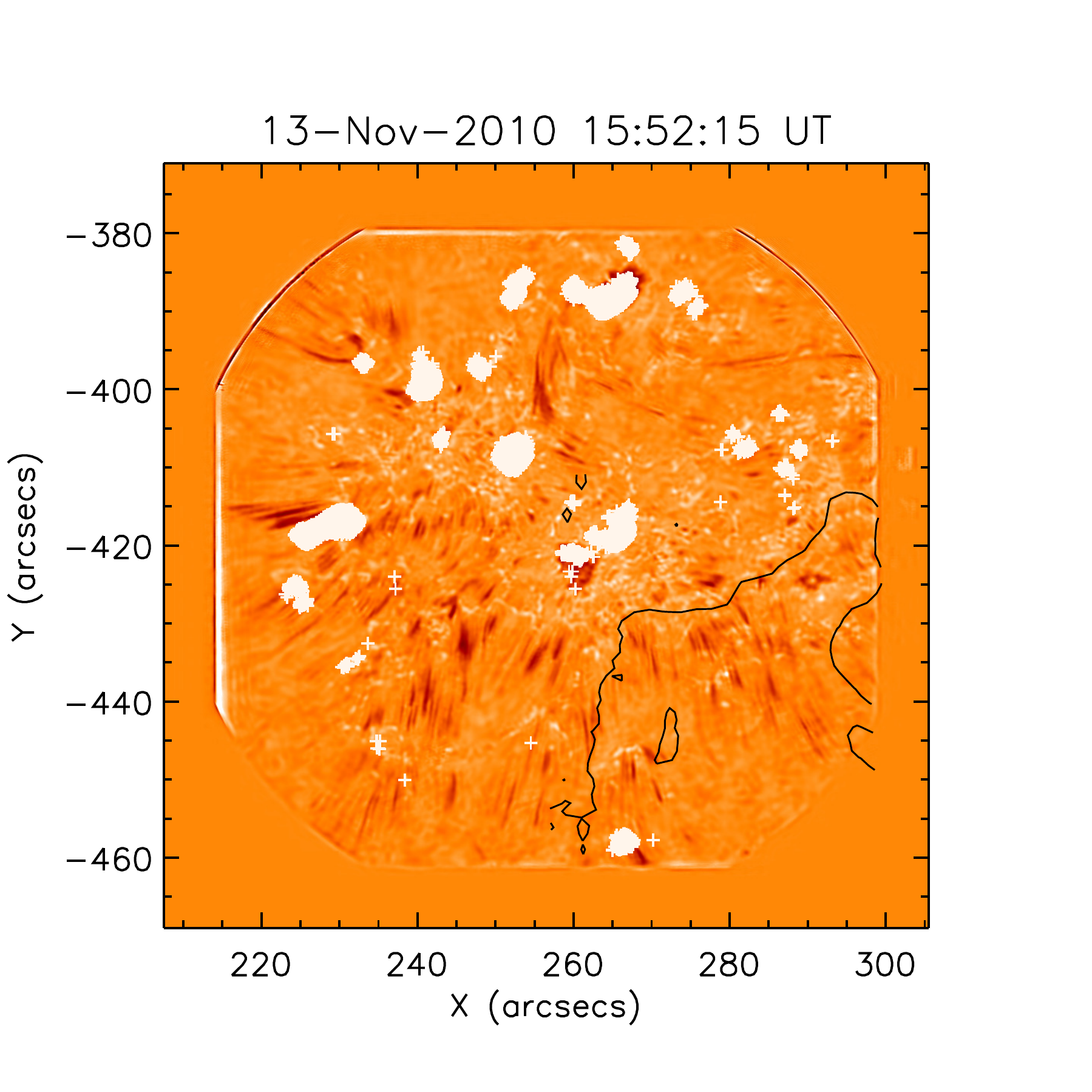}
\caption{IBIS image in the blue wing of the H$\alpha$ taken at 15:52\,UT with the contours of the upflow region superimposed in black. The white patches represent locations of blueward asymmetry detected in this dataset.}
\label{asymmetry}
\end{figure}

Blueward asymmetries have been detected near ARs in spectral lines with formation temperatures ranging from 0.6\,MK to 2\,MK at the locations of the upflows \citep{McIntosh2009}. Based on this  the authors made a conjecture that the mass carried up is originating from the lower layers of the solar atmosphere, i.e. the chromosphere. However, until now there has been no direct confirmation of the presence of such asymmetries in chromospheric lines. Here, we take advantage of the IBIS imaging spectrometer to check for  blue asymmetry in the H$\alpha$ line \citep[for details on the method used in this calculation refer to][]{Huang2014}. The outcome of our analysis is shown in Fig.~\ref{asymmetry} where we see that within the upflow region there are no noteworthy asymmetries. To reinforce our result we studied the asymmetries for the entire duration of the IBIS observations (see online material Fig.~\ref{fig_movie1}). In this movie it appears that there are some discontinuities in the number of asymmetry events detected from one frame to another. Although seeing effects cannot be completely ruled out to explain this variation, we believe that the dataset shown in this paper belongs to a time period of reasonable stable seeing. We attribute the frame-to-frame variation in the number of detected RBEs to the cadence of our dataset. A statistical study on the temporal evolution of RBEs by Sekse et al. (2013) has shown that their lifetime in H$\alpha$ data peaks at around 20\,s. This is shorter than the cadence of our dataset which is 24\,s. Therefore, many RBEs may be visible in only a single frame of our observations. There needs to be long duration asymmetries which are capable of sustaining several hours/days of coronal upflows. In the animation of Fig.~\ref{fig_movie1} (online material) we are able to show that the observed blue asymmetry does not last long enough even though a few frames show increased asymmetry within the upflow region.
 This result suggests that at chromospheric heights there are no activities which contribute to the coronal upflows in ARs. 

 \subsection{AIA}
\begin{figure}[ht!]
\centering
\vspace{-1cm}
\hspace{-1.3cm}
\includegraphics[width=10cm]{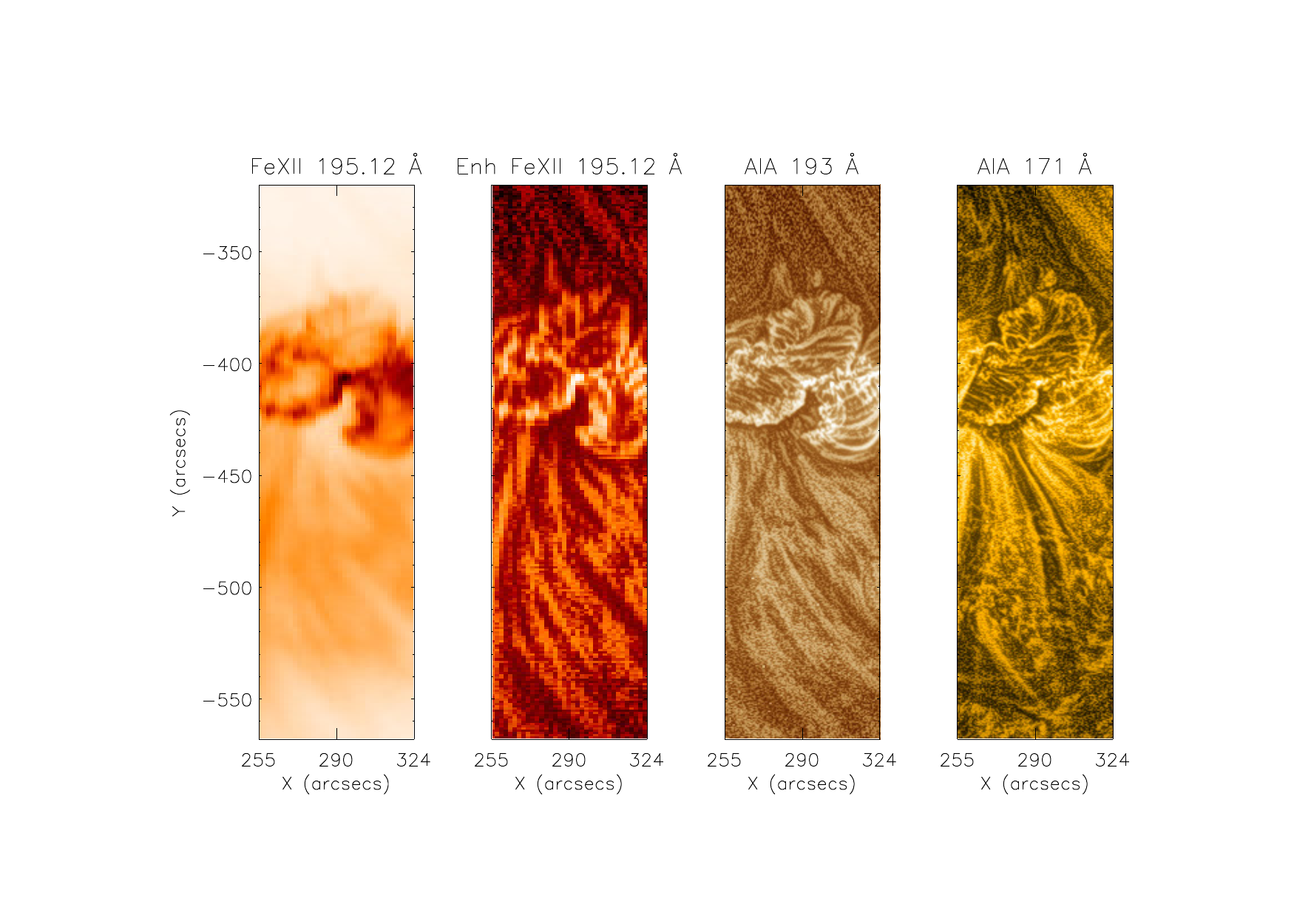}
\vspace{-1cm}
\caption{From left to right: EIS~Fe~{\sc xii}~195.12\,\AA\ image in inverse colour table, enhanced EIS~Fe~{\sc xii}~195.12\,\AA\ image, enhanced AIA~193\,\AA\ image and enhanced AIA~171\,\AA\  image taken at 15:00\,UT.}
\label{171_193}
\end{figure}

Combining imaging, spectroscopic  and magnetic field data of the solar atmosphere both in time and space is crucial for  understanding  the physical processes that generate  any  solar phenomenon. The combination is not straightforward for various reasons, e.g. longitudinal magnetic field data provide only one of the components of the magnetic field, imaging information often is contaminated by photons emitted from spectral lines with  a wide range of formation temperatures, spectral data have limited field of view and some require long exposures reducing the cadence of the data etc. These limitations can be challenged by using various image processing techniques to enhance weak signal, by applying magnetic field extrapolation, and combining  all the information for a better visual inspection.   

Several authors \citep{Warren2011, Young2012, DelZanna2011c} have raised the question on which coronal imaging channel best represents the coronal structures that display plasma upflows in the EIS Fe\,{\sc xii} lines.  The reason for this is the contribution of a wide range of spectral lines to the wavelength range covered by the AIA 171\,\AA\ and 193\,\AA, and TRACE 171\,\AA\ channels. The 171\,\AA\  channel is dominated by
Fe\,{\sc ix} but if a feature at low temperature is observed, the recorded emission will be at temperatures log\,T (K) $<$
5.7 \citep{Brooks2011, Vanninathan2012}. The channel also has a contribution from Fe\,{\sc x}. The AIA 193\,\AA\ channel is dominated by three
Fe\,{\sc xii} lines but it also has a significant contribution from the transition region from unidentified lines \citep{DelZanna2011c}.

\begin{figure*}[ht!]
\centering
\hspace{-1.3cm}
\includegraphics[width=6cm]{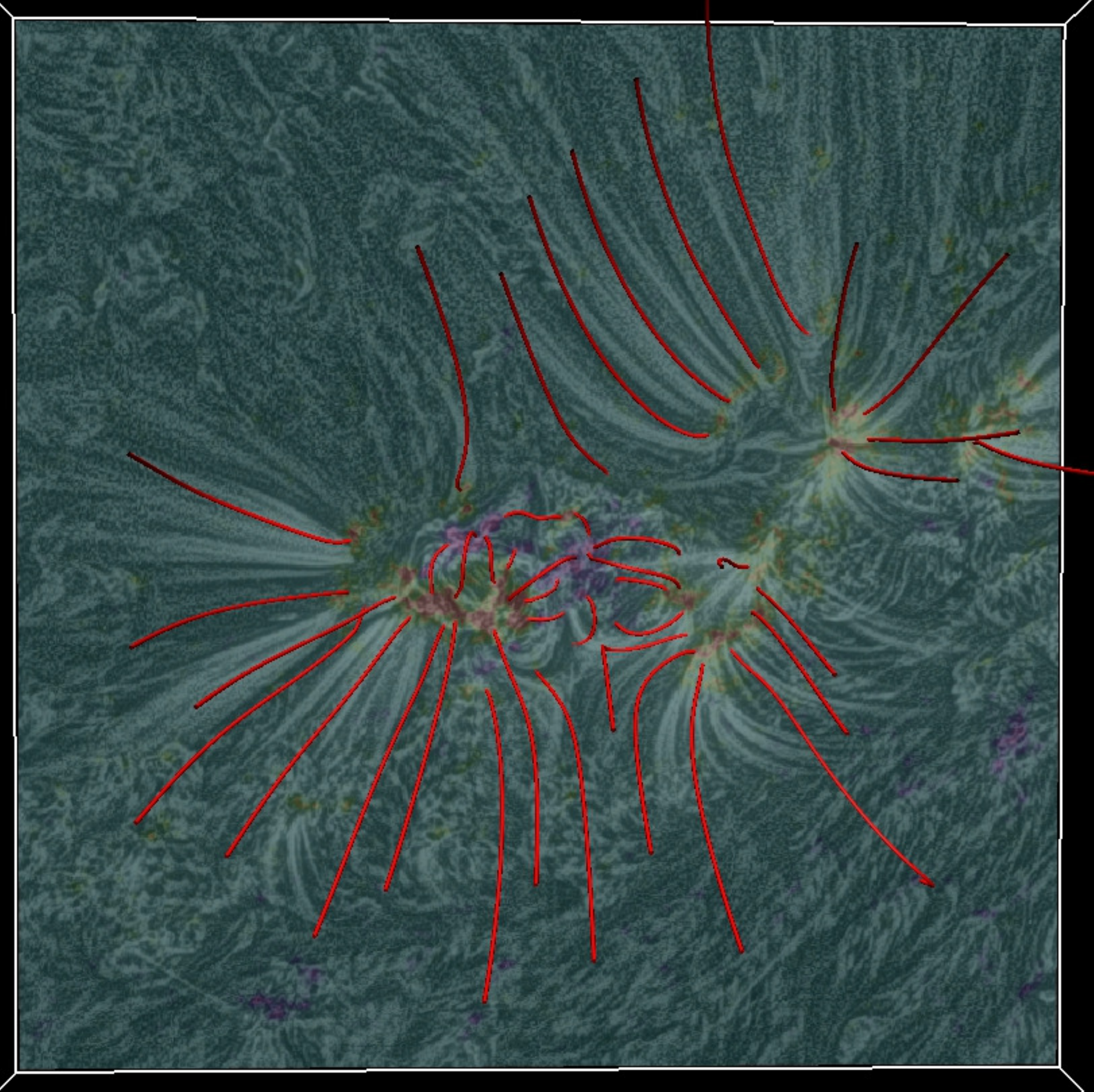}
\hspace{1cm}
\includegraphics[width=6cm]{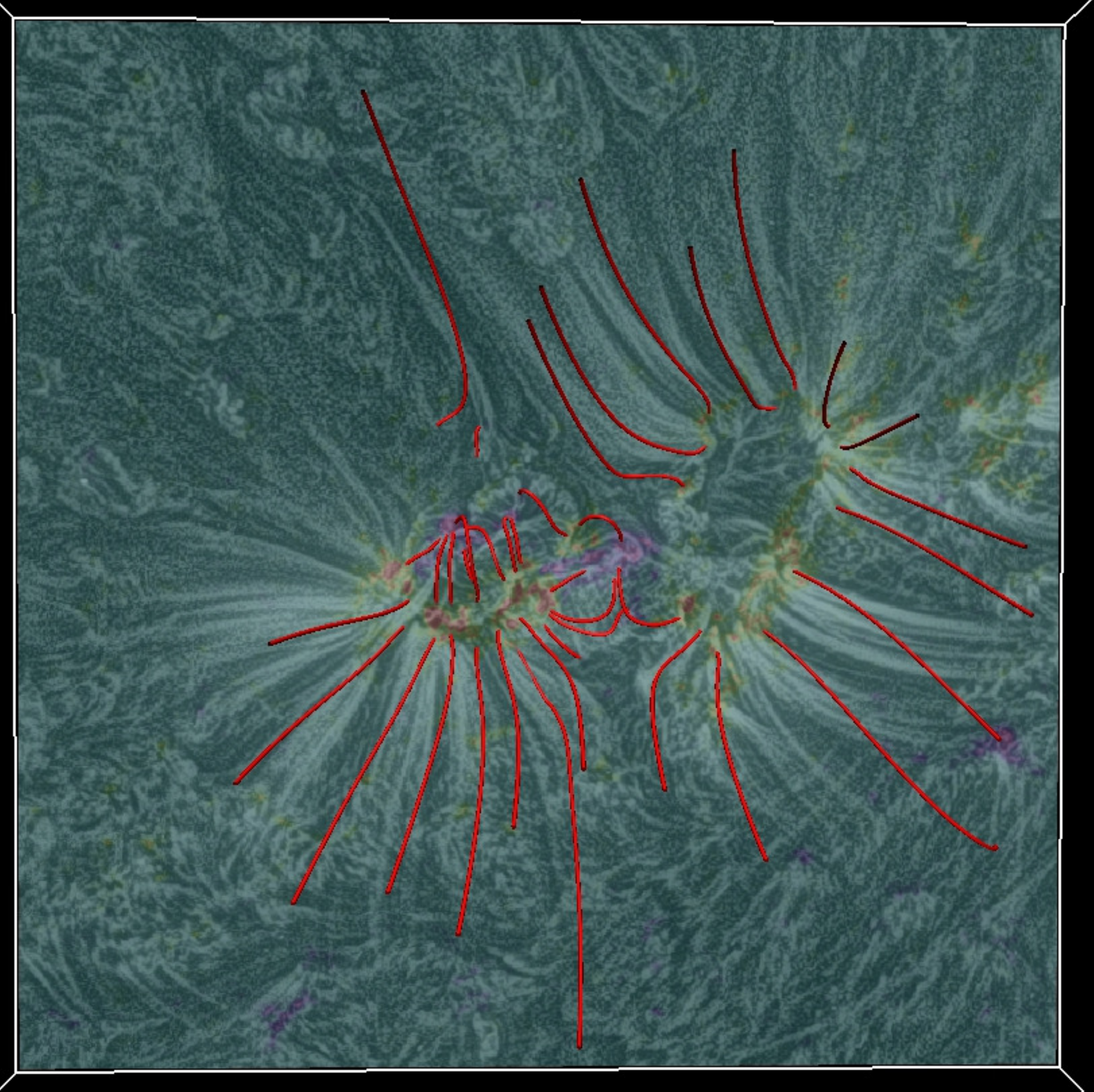}
\caption{Enhanced AIA~171\,\AA\ images taken at 18:00\,UT on November 12 \textbf{(left)} and 16:00\,UT on November13, 2010 \textbf{(right)} together with the magnetic field lines obtained from potential field extrapolation.}
\label{aia_extrap}
\end{figure*}

We made a  comparison of the upflow region in MGN  enhanced AIA 171\,\AA\ and 193\,\AA\ images, and EIS Fe\,{\sc xii} intensity image shown in Fig.~\ref{171_193}.  Clearly the upflow occurs outside the existing closed loop system with  a number of quasi-open loops expanding out from the edge of the AR.  To verify this and for the purpose of the follow-up simulations, we produced a potential field model given in Fig.~\ref{aia_extrap}. From the visual inspection of panels 2, 3  and 4 (from left to right) in Fig.~\ref{171_193}, it is evident that most of the loop structures detected in EIS Fe\,{\sc xii} correspond to the loops seen in AIA 193\,\AA. The AIA 171\,\AA\ image shows many of the EIS Fe\,{\sc xii} loops but not all. At the same time several loops  seen in AIA 171\,\AA\ are not observed in the  EIS Fe\,{\sc xii} and AIA 193\,\AA\ images.  Thus, we conclude that investigating the upflow feature in  EIS Fe\,{\sc xii} and AIA 193\,\AA\ and to some extent in AIA 171\,\AA\ can be done with confidence. 

In the low temperature lines like Si\,{\sc vii} and Fe\,{\sc viii},  bright fan-like structures are observed  dominated
by downflows. \citet{Warren2011}  concluded that the morphology of the cooler fan and the upflow structure is different. The upflows appear to be 
observed in regions where there is no emission in the cooler lines. \citet{Ugarte-Urra2011} confirmed this result and discussed the idea of two populations of loops, open and closed, with blue-shifted emission seen in the open and red-shifted in the closed loops. \citet{Young2012} examined in great detail the `filamentary features' observed in the cooler EIS Si\,{\sc vii} and hotter Fe\,{\sc x} lines and suggested a possible scenario of multi-strand structure not resolved by  EIS but detectable in TRACE imaging data. For a two strand example, one strand is stationary and emitting in Fe\,{\sc x} along a large portion of its length, while along a second strand  the plasma is downflowing and emitting in Fe\,{\sc viii}.
 The Fe\,{\sc x} strand is located higher than the Fe\,{\sc viii} strand and since the Fe\,{\sc x} strand must cool before descending
into the photosphere, it too must emit in Fe\,{\sc viii}, but this emission will be much more compact than the emission coming from
the Fe\,{\sc viii} strand. For more details on this scenario please see \cite{Young2012}.
 Red-shifted emission in Si\,{\sc vii} and Fe\,{\sc viii} and blue-shifted emission in Fe\,{\sc xii}  are reported in all AR upflow papers so far. 
The same picture can be seen in all AR data found in the solar archive that display `outflow'. For the present case we have found a second velocity component at $\sim$105\,\kms. A very plausible scenario is that in one EIS spatial pixel and even one AIA pixel, there exists as suggested by  \cite{Young2012} several strands along which plasma at different speeds is ejected. The two outflow velocity components suggest that possibly these two different flows are produced by two different physical mechanisms, e.g. magnetic reconnection  and pressure-driven. This implies   that we need higher spatial resolution both spectral and imaging observations to resolve this question. 

\section{Conclusions and summary}
The abundance of studies on AR upflows demonstrates the importance of the subject for solar and interplanetary physics. In the present study we have used multi-instrument data to investigate the origin and physical properties of AR upflows. We have studied an AR observed  on 2010 November 13 using spectral and imaging data from the entire solar atmosphere including photospheric magnetic field. The basic characteristics of the AR upflows we have studied here, Doppler shift and electron density, are typical to those reported in prior studies. We observed blue-shifted velocities of the order of 5 -- 20\,\kms\ and densities around 1.8$\times10^9$\,cm$^{-3}$ at 1\,MK within the upflow region. The time variation of the electron density does not show any significant
change within 3$\sigma$ errors suggesting a physical mechanism that can provide a smooth continuous flow. The spectral lines formed at 1\,MK (Fe\,{\sc xii} and Fe\,{\sc xiii}) showed an enhanced blue wing component with velocities around 105\,\kms. This blue wing asymmetry was only present close to the footpoints of the upflow region and was persistent during the entire observing period with variations of the order of $\pm$15\,\kms. By studying the magnetic flux evolution within the upflow region, we have found that in the current dataset flux emergence does not take place but rather general flux diffusion is responsible for the coronal upflows.

Many observations and models (as mentioned in Section 1) support the theory that the chromosphere plays an important role in the occurrence of AR upflows with chromospheric jets as the drivers of these upflows. For the first time, using imaging spectroscopy of the chromosphere in the H$\alpha$ line, we have suggested that this region of the solar atmosphere is not where coronal upflows originate. A detailed study of  jet-like chromospheric  phenomena show plasma ejected upwards and returning back to its origin after a short interval of time. This implies that these are not type~II spicules or RBEs which are believed to sustain an AR upflow. In addition, we measured the blueward asymmetry in the H$\alpha$ line, which is considered a characteristic of RBEs \citep{Rouppe2009}, and found no such asymmetry within the upflow region. These results indicate that  chromospheric jets do not contribute to AR upflows.

We should also remark on the recently discovered small-scale jet-like features in the transition region \citep{Tian2014}. Suitable observations (e.g. IRIS/EIS co-observations in AR upflow region) in the future will have the opportunity to investigate whether these features can be regarded as a possible contributor to  AR upflows.

We propose that convective motions drive the magnetic fields which are then responsible for the observed AR upflows. This idea is further investigated through data-driven modelling in Paper~II, even though this particular model fails to reproduce the observed upflows.

\begin{acknowledgements} 
The authors are immensely grateful to Huw Morgan for providing us with the MGN code. KV and MM would like to thank Kanzelh{\" o}he Observatory for letting us use their facility for discussions while completing the paper. Particular thanks to Werner P{\" o}tzi for his hospitality during our stay at Kanzelh{\" o}he Observatory. IBIS data reduction was accomplished with help from Kevin Reardon. KV acknowledges the Austrian Science Fund (FWF): P24092-N16. MM is funded by the Leverhulme trust. ZH is supported by the China 973 program (2012CB825601) and the National Natural Science Foundation of China under contract 41404135. Research at Armagh Observatory is grant-aided by the N. Ireland Department of Culture, Arts and Leisure and via grants ST/J001082/1 and ST/M000834/1 from the UK STFC.
Hinode is a Japanese mission developed and launched by ISAS/JAXA, with NAOJ as domestic partner and NASA and STFC (UK) as international partners.
 CHIANTI is a collaborative project involving NRL (USA), RAL (UK), and the Universities: College London (UK), of Cambridge (UK), George Mason (USA), and of Florence (Italy).
The AIA and HMI data are courtesy of SDO (NASA). The NSO is operated by the Association of Universities for Research in Astronomy, Inc., for
the National Science Foundation. IBIS was built by the INAF/Osservatorio Astrofisico di Arcetri with contributions from the Universities
of Firenze and Roma Tor Vergata, the National Solar Observatory, and the Italian Ministries of Research and Foreign Affairs. VAPOR software Imagery produced by VAPOR (www.vapor.ucar.edu), a product of the Computational Information Systems Laboratory at the National Center for Atmospheric Research.
\end{acknowledgements}

\bibliographystyle{aa}
\bibliography{Vanninathan}

\begin{appendix}
\section{Online material}
\begin{figure}[!h]
\centering
\includegraphics[width=8cm]{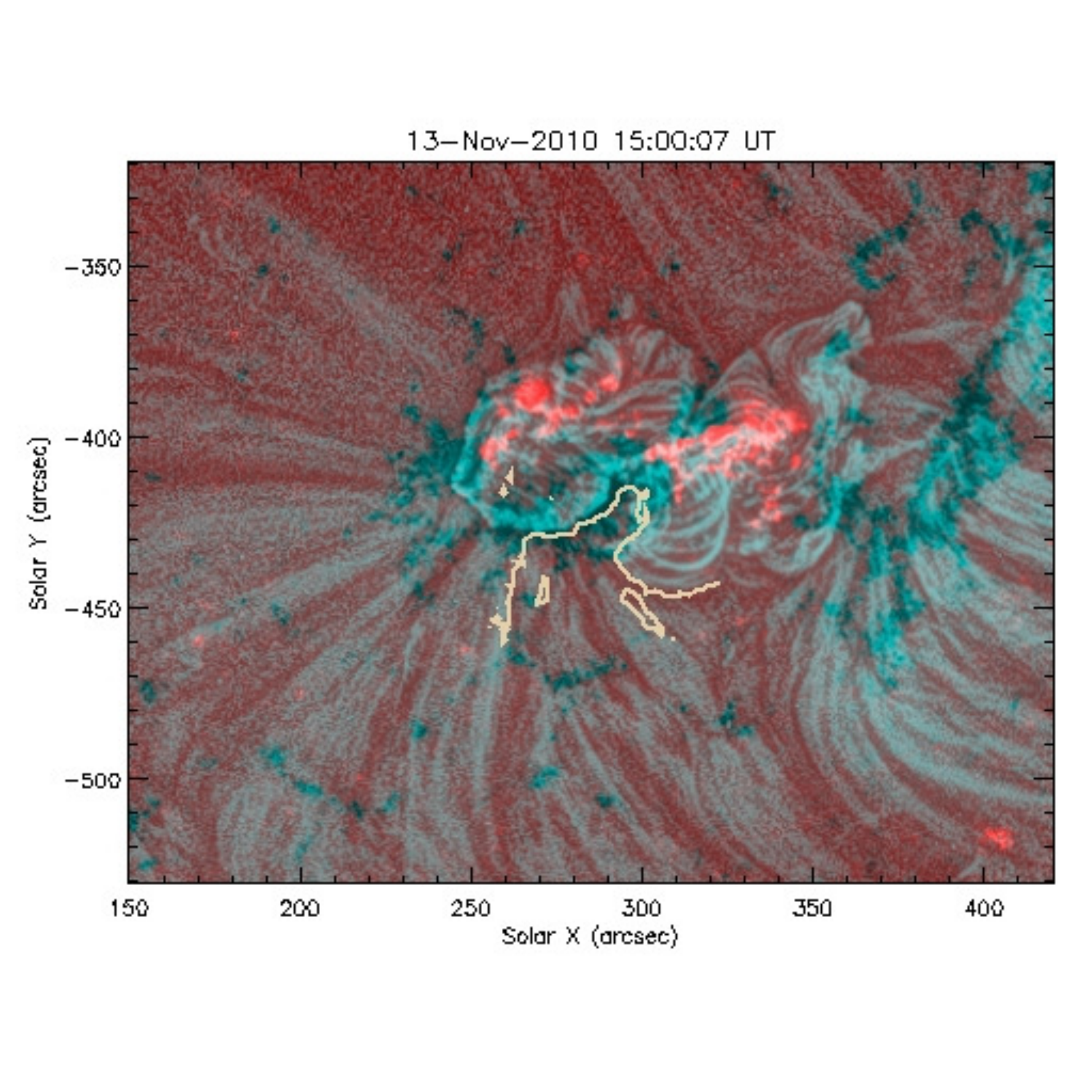}
\caption{Animation showing the HMI  magnetic field images in the background and the loops as seen in AIA 193\,\AA\ images. The contour of the upflow region is overplotted with yellow solid line.}
\label{fig_movie2}
\end{figure}

\begin{figure}[!h]
\centering
\includegraphics[width=8cm]{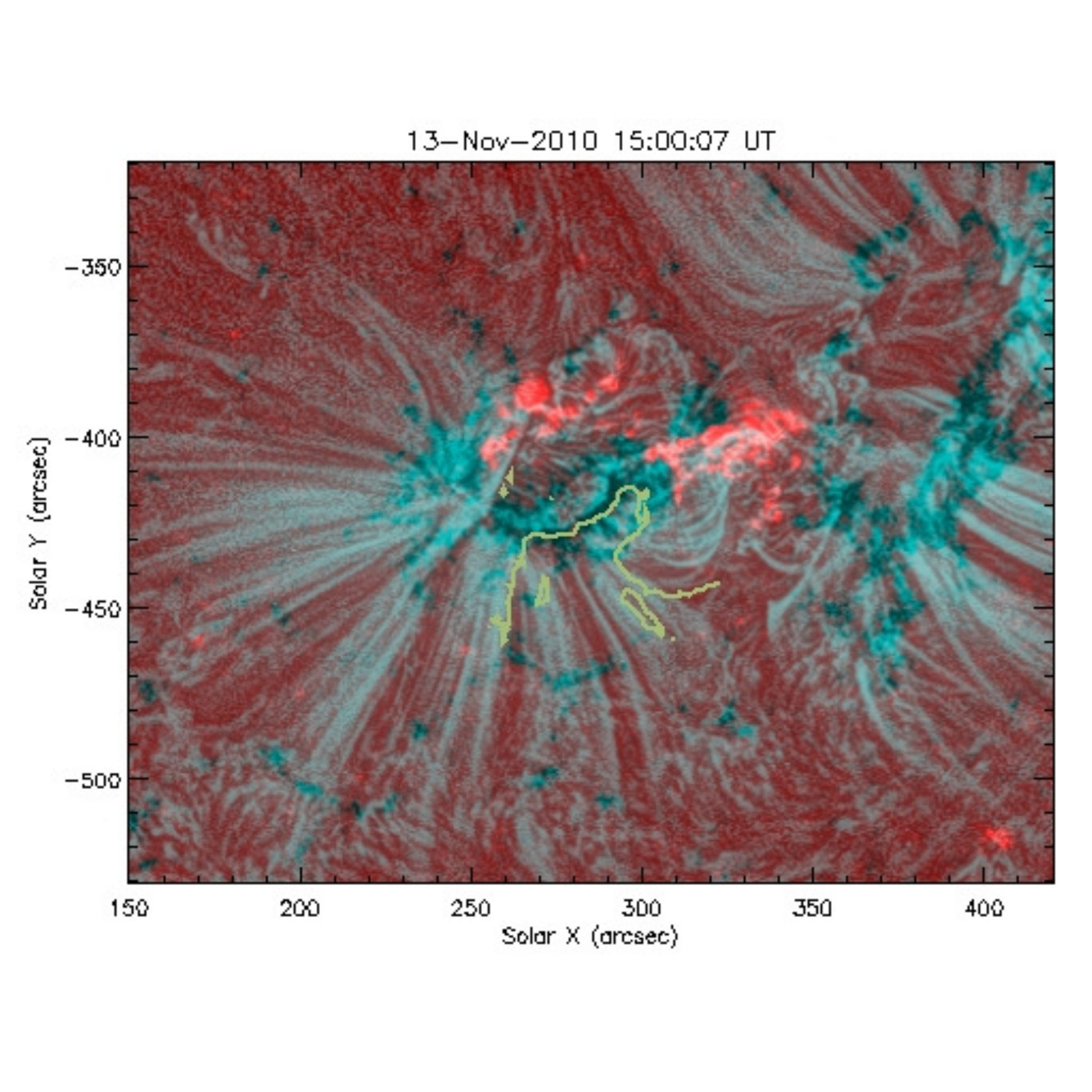}
\caption{Animation showing the HMI  magnetic field images in the background and the loops as seen in AIA 171\,\AA\ images. The contour of the upflow region is overplotted with green solid line.}
\label{fig_movie3}
\end{figure}

\begin{figure}[!h]
\centering
\includegraphics[width=8cm]{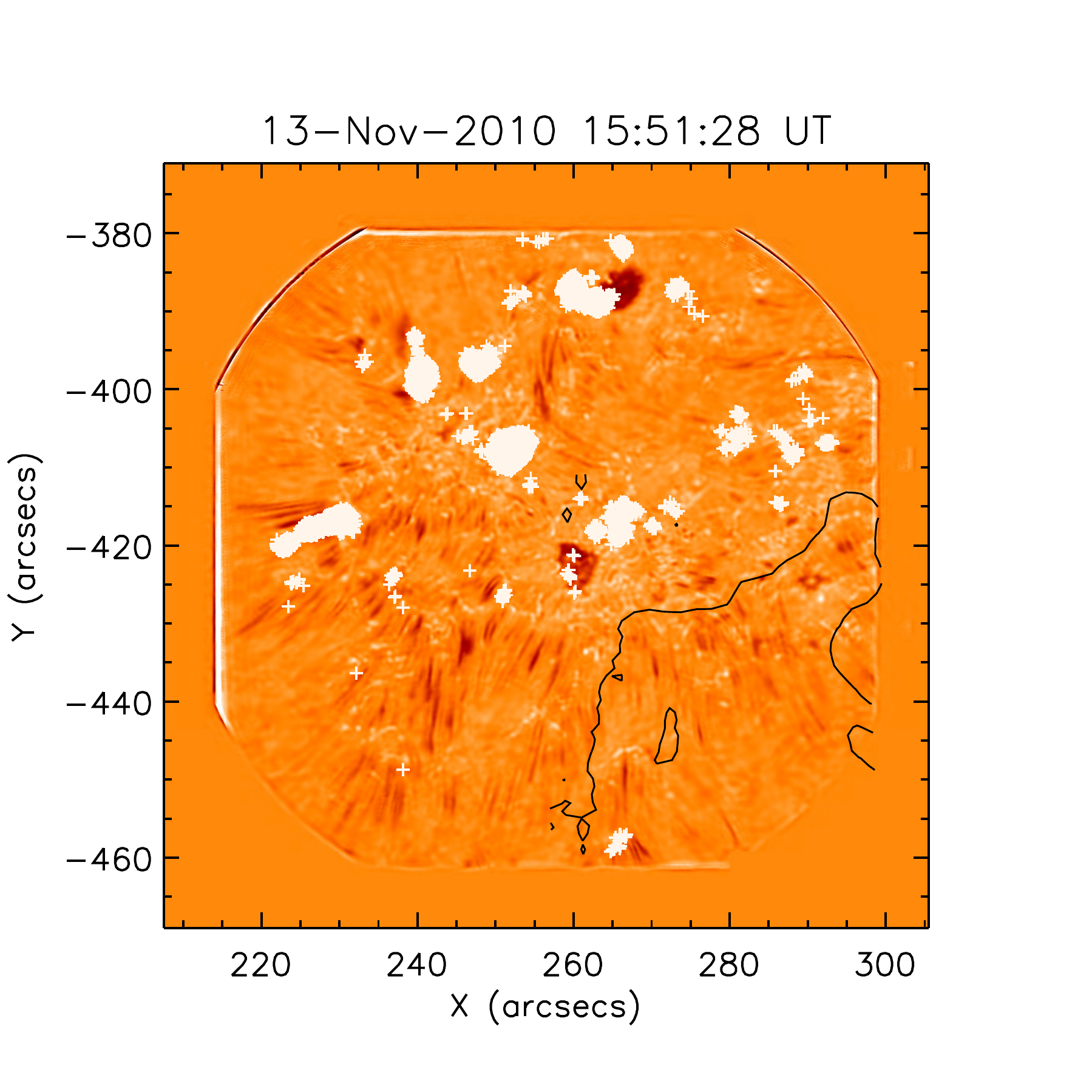}
\caption{Animation of IBIS images taken in the blue wing of the H$\alpha$ line with the contour of the upflow region superimposed in black. The white patches represent locations of blueward asymmetry detected in this dataset.}
\label{fig_movie1}
\end{figure}

\end{appendix}

\end{document}